\documentclass{rsproca}%%%%where rsproca is the template name

\newcommand{\V}[1]{\mathbf{#1}} 
\newcommand{\T}[1]{\texttt{#1}} 
\newcommand\Alfven{Alfv\'en }
\newcommand\Alfvenic{Alfv\'enic }
\newcommand{\figref}[1]{Fig.~\ref{#1}}
\newcommand{\secref}[1]{\S\ref{#1}}
\newcommand{\xhat}{\mbox{$\hat{\mathbf{x}}$}} 
\newcommand{\yhat}{\mbox{$\hat{\mathbf{y}}$}} 
\newcommand{\zhat}{\mbox{$\hat{\mathbf{z}}$}}

\newcommand{\XCC}{\mbox{$C(\delta n, \delta B_{\parallel})~ $}}

\usepackage{color}

\begin{document}

%%%% Article title to be placed here
%\title{A Self-Consistent Picture of the Dynamics and Dissipation of Plasma Turbulence}
\title{A Dynamical Model of Plasma Turbulence in the Solar Wind}

\author{%%%% Author details
G. G. Howes}

%%%%%%%%% Insert author address here
\address{Department of Physics and Astronomy, University of
Iowa, Iowa City, IA 52242, USA }
%%%% Subject entries to be placed here %%%%
\subject{Plasma Physics, Astrophysics, Solar System}

%%%% Keyword entries to be placed here %%%%
\keywords{Plasma Turbulence, Solar Wind, Alfven Waves, Plasma Kinetic Theory, Turbulent Dissipation, Plasma Heating}

%%%% Insert corresponding author and its email address}
\corres{Gregory G. Howes\\
\email{gregory-howes@uiowa.edu}}

%%%% Abstract text to be placed here %%%%%%%%%%%%
\begin{abstract}
A dynamical approach, rather than the usual statistical approach, is
taken to explore the physical mechanisms underlying the nonlinear
transfer of energy, the damping of the turbulent fluctuations, and the
development of coherent structures in kinetic plasma turbulence.  It
is argued that the linear and nonlinear dynamics of \Alfven waves are
responsible, at a very fundamental level, for some of the key
qualitative features of plasma turbulence that distinguish it from
hydrodynamic turbulence, including the anisotropic cascade of energy
and the development of current sheets at small scales. The first
dynamical model of kinetic turbulence in the weakly collisional solar
wind plasma that combines self-consistently the physics of \Alfven
waves with the development of small-scale current sheets is presented
and its physical implications are discussed. This model leads to a
simplified perspective on the nature of turbulence in a weakly
collisional plasma: the nonlinear interactions responsible for the
turbulent cascade of energy and the formation of current sheets are
essentially fluid in nature, while the collisionless damping of the
turbulent fluctuations and the energy injection by kinetic
instabilities are essentially kinetic in nature.
\end{abstract}
%%%%%%%%%%%%%%%%%%%%%%%%%%%

%%%%%%%%%% Insert the texts which can accomdate on firstpage in the tag "fmtext" %%%%%

\begin{fmtext}
%===================================================================
\section{Introduction}

Five centuries ago, Leonardo da Vinci first marveled at the intricate
flow patterns arising in a turbulent cascade of water.  But the first
quantitative understanding of the physics of hydrodynamic turbulence
had to wait until the seminal work by Kolmogorov in the 1940s, and
humankind's grasp of hydrodynamic turbulence remains yet incomplete.
To advance the frontiers of our knowledge beyond the limits of our
terrestrial environment, we strive to understand the impact of
turbulence on the astrophysical plasmas that comprise most of the
visible matter in the universe.

On the face of it, it would appear that unraveling the details of
plasma turbulence, with electromagnetic forces in addition to the gas
pressure force that occurs
\end{fmtext}

%%%%%%%%%%%%%%% End of first page %%%%%%%%%%%%%%%%%%%%%

\maketitle

\noindent in hydrodynamics, would be a much more challenging task than the study
of turbulence in water and air.  But the study of plasma turbulence is
facilitated by one fundamental difference between a hydrodynamic fluid
and a plasma: in a magnetized plasma, magnetic tension provides a
restoring force that is absent in hydrodynamics. The importance of
this distinction cannot be overstated, because magnetic tension
supports the propagation of a fundamental wave mode, the \Alfven wave,
that plays a governing role in plasma turbulence.

The linear \Alfvenic response, due to magnetic tension, of the plasma
to applied perturbations provides a critical theoretical foothold in
the quest to understand the dynamics and energetics of plasma
turbulence. Specifically, in the limit of weak
nonlinearity,\footnote{Weak nonlinearity means that the magnitude of
  the nonlinear terms in the equations of evolution are small relative
  to the magnitude of the linear term that supports the propagation of
  \Alfven waves.} it is possible to derive rigorous asymptotic
analytical solutions of the nonlinear dynamics. It is precisely this
nonlinear physics that underlies the turbulent cascade of energy from
large to small scales, one of the most important impacts of turbulence
on astrophysical environments. In practice, such rigorous solutions in
the limit of weak nonlinearity can often be pushed to the limits of
strong turbulence while remaining at least qualitatively correct, even
though such solutions are formally well beyond their regime of
applicability.

Lacking this foothold of linear physics, hydrodynamics ruthlessly
forces the physicist to resort almost immediately to a statistical
approach to describe the turbulent evolution, with the typical focus on
the power spectra, the development of intermittency, and the scaling
of higher-order statistics.  In the weakly collisional solar
wind, researchers at the forefront strive to illuminate the physics of
the nonlinear energy transfer to small scales, the kinetic mechanisms
of dissipation of the turbulent fluctuations, and the resulting plasma
heating.  To illuminate these mechanisms, I contend that we must step
beyond the usual statistical treatments, adopting instead a dynamical
approach, to determine definitively the dominant physical processes at
play in the evolution of plasma turbulence. Only with a dynamical
understanding of both the nonlinear wave-wave interactions responsible
for the turbulent cascade and the collisionless wave-particle
interactions responsible for the dissipation will it be possible to
construct a predictive theory of plasma turbulence and its effect on
energy transport and plasma heating.

A significant fraction of the heliospheric turbulence research
community appears to believe, likely based on the analogy with
hydrodynamics, that linear physics properties are not relevant to
strong plasma turbulence.  On the contrary, there are numerous
counterexamples in which linear physics properties have been shown to
be relevant to strong plasma turbulence
\cite{Maron:2001,Cho:2003,Alexandrova:2008,Howes:2008a,Howes:2008b,Svidzinski:2009,Sahraoui:2010b,Hunana:2011,Howes:2011a,TenBarge:2012b,Chen:2013a}
and in some cases significant insights into the nature of magnetized
plasma turbulence have been achieved by the exploitation of intuition
from the linear physics properties
\cite{Howes:2011c,Howes:2012a,Klein:2012,Salem:2012,Chen:2013b,Howes:2013a,Nielson:2013a,Klein:2014a}.
One particularly significant example is that the qualitative picture
of the nonlinear energy transfer in the weak turbulence limit persists
in the limit of strong turbulence \cite{Howes:2014d}.  In fact, there
is a long history of linear wave properties being explicitly or
implicitly used to analyze spacecraft measurements and numerical
simulations of plasma turbulence
\cite{Coleman:1968,Belcher:1971,Tu:1984,Matthaeus:1990,Matthaeus:1994,
  Tu:1994,Verma:1995,Leamon:1998b,Quataert:1998,Leamon:1999,Quataert:1999,Leamon:2000,Stawicki:2001,Verma:2004,Bale:2005,Markovskii:2006,Hamilton:2008,Howes:2008b,Sahraoui:2009,Schekochihin:2009,Chandran:2010a,Chen:2010b,Howes:2010d,Podesta:2010a,Sahraoui:2010b,Saito:2010,Rudakov:2012}.

In fact, as will be explained below, the linear and nonlinear dynamics
of \Alfven waves are responsible, at a very fundamental level, for
some of the key qualitative features of plasma turbulence that
distinguish it from hydrodynamic turbulence, including the anisotropic
cascade of energy and the development of current sheets at small
scales. In the opinion of this author, to attempt a purely statistical
investigation of plasma turbulence, analogous to the common approach
in hydrodynamic turbulence, without exploiting the plasma physics of
the turbulent fluctuations is bound to meet with moderate success at
best.

The current frontier of research on plasma turbulence in the weakly
collisional solar wind is framed by three fundamental questions:
\begin{enumerate}
\item {\bf What is the physical mechanism underlying the nonlinear
  transfer of energy from large to small scales?} This question must
  be answered both for the MHD regime of the inertial range and the
  kinetic regime of the dissipation range.\footnote{I denote here the
    MHD regime as $k_\perp \rho_i \ll 1$ and the kinetic regime as
    $k_\perp \rho_i \gtrsim 1$, although compressible fluctuations in
    the weakly collisional solar wind require a kinetic treatment,
    even at the large scales of the inertial range
    \cite{Schekochihin:2009,Howes:2009b,Klein:2012}. Here $k_\perp$
    refers to the component of the wavevector perpendicular to the
    local mean magnetic field $\V{B}_0$ and $\rho_i = v_{ti}/\Omega_i$
    is the thermal ion Larmor radius, where the ion thermal velocity
    is defined by $v_{ti}^2= 2T_i/m_i$ (with temperature expressed in
    units of energy) and the ion cyclotron frequency is given by
    $\Omega_i = q_i B_0/(m_i c)$.}
\item {\bf What are the dominant physical mechanisms responsible for
  damping the turbulent electromagnetic fluctuations?} Note that which
  mechanisms dominate may depend on both the plasma parameters and the
  characteristics of the turbulence, such as the scale and amplitude
  of the energy injection. These mechanisms determine the partitioning
  of dissipated turbulent energy into heat, or other energization, of
  the protons, electrons, and minor ions.
\item {\bf How do coherent structures arise from and/or affect both
  the nonlinear energy transfer and the dissipation mechanism?} The
  concentration of dissipation in coherent structures, specifically
  current sheets, is well established by numerical simulations of
  plasma turbulence and is inferred from spacecraft measurements in
  the solar wind. 
\end{enumerate}
The ultimate goal of the study of space and astrophysical plasma
turbulence is to develop a sufficiently detailed understanding of the
physics to enable the construction of predictive models of the
nonlinear energy transfer, the damping of the turbulent fluctuations,
and the resulting heating of the plasma species, for example the
heating of the solar corona \cite{Cranmer:2015}. In addition to
establishing a more thorough knowledge of the physics of the
heliosphere---our home in the universe---it will provide a crucial
foundation for an improved understanding of complex astrophysical
phenomena in remote regions of the cosmos.

Below I present the first dynamical model of kinetic turbulence in
the weakly collisional solar wind plasma that combines
self-consistently the physics of \Alfven waves with the development of
small-scale current sheets. Based on this model, I will address each
of the questions above, as well as numerous subsidiary issues, as we
follow the turbulent cascade of energy from large to small scales in
the solar wind.

%===================================================================
\section{The Dynamics of the Turbulent Cascade}
Incompressible MHD is one of the most simple, self-consistent
descriptions of turbulent plasma dynamics.  Although incompressible
MHD lacks much of the rich physical behavior possible in the weakly
collisional plasma conditions relevant to the solar wind, I argue here
that incompressible MHD systems appear to contain the minimum number
of physical ingredients necessary to yield the key qualitative
features of plasma turbulence that distinguish it from hydrodynamic
turbulence, specifically the anisotropic cascade of energy and
dissipation dominantly occurring within current sheet
structures. These key features persist as more physically complete
plasma descriptions are adopted, yet do not occur in hydrodynamic
turbulence.

Early research on incompressible MHD turbulence in
the 1960s \cite{Iroshnikov:1963,Kraichnan:1965} suggested that
nonlinear interactions between counterpropagating \Alfven waves---or
\Alfven wave collisions---support the turbulent cascade of energy from
large to small scales. The incompressible MHD equations, expressed
here in the symmetrized Elsasser form \cite{Elsasser:1950}, are
\begin{equation}
\frac{\partial \V{z}^{\pm}}{\partial t} 
\mp \V{v}_A \cdot \nabla \V{z}^{\pm} 
=-  \V{z}^{\mp}\cdot \nabla \V{z}^{\pm} -\nabla P/\rho_0,
\label{eq:elsasserpm}
\end{equation}
and $\nabla\cdot \V{z}^{\pm}=0$. Here $\V{v}_A =\V{B}_0/\sqrt{4
  \pi\rho_0}$ is the \Alfven velocity due to the local mean magnetic
field $\V{B}_0=B_0 \zhat$ where the total magnetic field is
$\V{B}=\V{B}_0+ \delta \V{B} $.  Note that the direction of the local
mean magnetic field plays a key role in the physics of incompressible
MHD turbulence \cite{Cho:2000}, and here the coordinate system is
chosen so that the unit vector direction $\zhat$ is aligned with the
mean field direction.  $P$ is total pressure (thermal plus magnetic),
$\rho_0$ is mass density, and $\V{z}^{\pm} = \V{u} \pm \delta
\V{B}/\sqrt{4 \pi \rho_0}$ are the Elsasser fields which represent
waves that propagate up or down the mean magnetic field. The
$\V{z}^{\mp}\cdot \nabla \V{z}^{\pm} $ term governs the nonlinear
interactions between counterpropagating \Alfven waves, denoted
\emph{\Alfven wave collisions}.

The strength of the nonlinearity in incompressible MHD turbulence may
be characterized by the \emph{nonlinearity parameter}, $\chi \equiv
|\V{z}^{-}\cdot \nabla \V{z}^{+}|/|\V{v}_A \cdot \nabla \V{z}^{+}|$,
the ratio of the magnitude of the nonlinear term to that of the linear
term in \eqref{eq:elsasserpm}. Weak incompressible MHD turbulence
corresponds to the limit $\chi \ll 1$, while a state of strong
incompressible MHD turbulence is characterized by $\chi \sim 1$
\cite{Sridhar:1994,Goldreich:1995}.

%===================================================================
\subsection{How is Energy Transferred to Small Scales?}
\label{sec:cascade}
\begin{figure}[t]
\vbox{\hsize \textwidth \hspace*{0.15in}(a)\hspace{2.1in}  (b)}
\vspace{-0.15in}
\centering\hbox{\resizebox{!}{2.3in}{\includegraphics*[0.32in,2.25in][8.2in,9.75in]{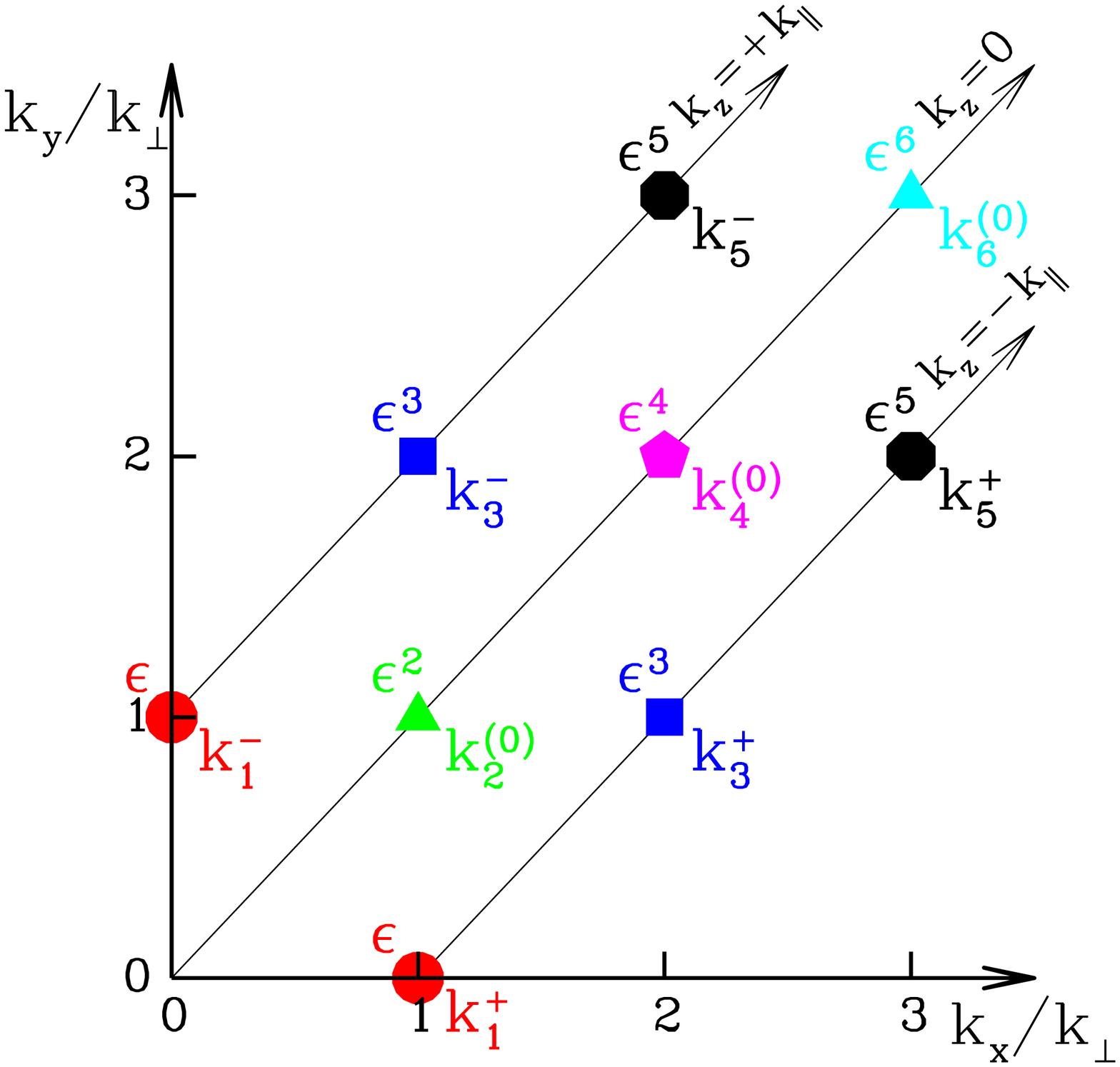}} \hfill 
\resizebox{!}{2.3in}{\includegraphics{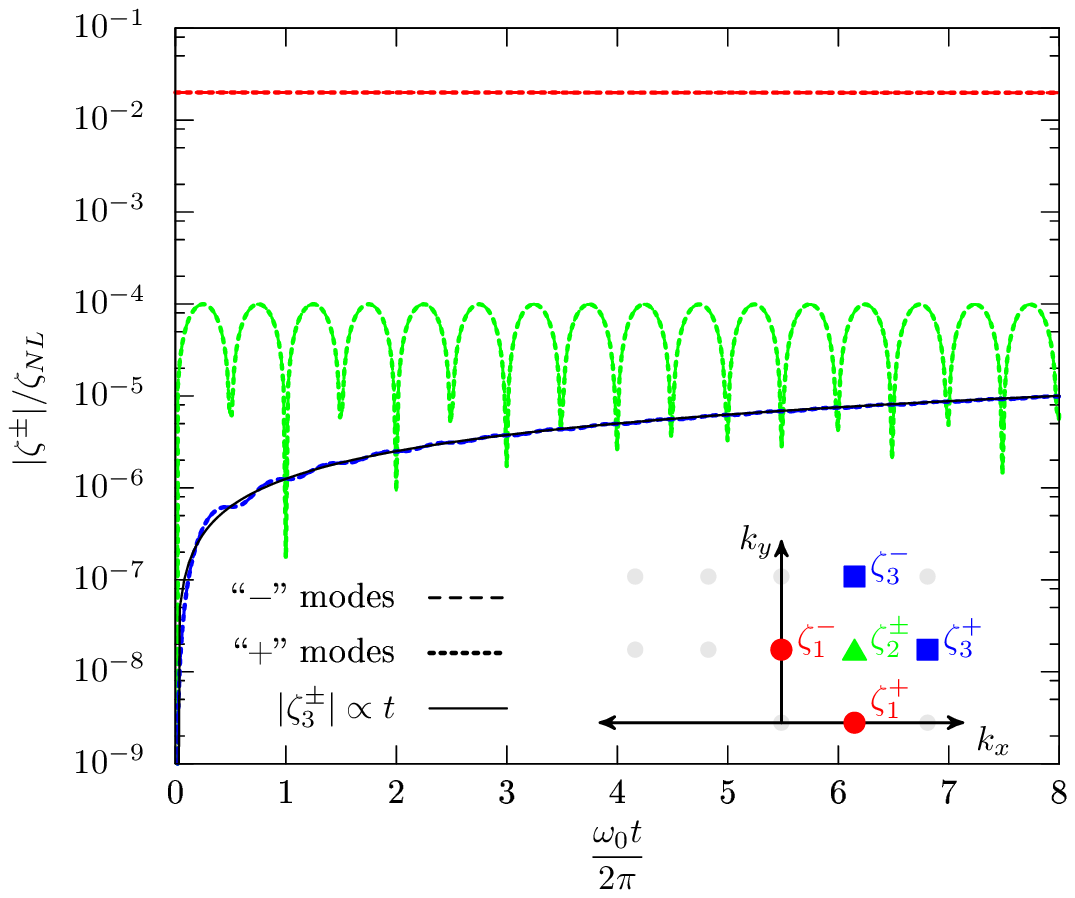}}}
\caption{(a) Perpendicular Fourier modes nonlinearly generated by an
  \Alfven wave collision between counterpropagating parent \Alfven
  waves $\V{k}_1^+$ and $\V{k}_1^-$. From Howes 2014
  \cite{Howes:2014d}. (b) In the weakly nonlinear limit, evolution of
  the normalized amplitude $|\zeta^\pm|/\zeta_{NL}$ of the key Fourier
  modes involved in the nonlinear energy transfer vs.~time $\omega_0
  t/ 2 \pi$ over eight periods of the primary \Alfven waves. Here
  $\V{z}^{\pm} = \zhat \times \nabla_\perp \zeta^{\pm}$, $\zeta_{NL}
  \equiv k_\parallel v_A/k_\perp^2$, and $\omega_0=k_\parallel v_A$.
  From Nielson, Howes, and Dorland 2013
  \cite{Nielson:2013a}\vspace*{-10pt} }
\label{fig:modes_o5}
\end{figure}

%\vspace*{-10pt}

Following significant previous studies on weak incompressible MHD
turbulence \cite{Sridhar:1994,Montgomery:1995,Ng:1996,Galtier:2000},
the nonlinear energy transfer in \Alfven wave collisions has recently
been solved analytically in the weakly nonlinear limit, $\chi \ll 1$
\cite{Howes:2013a}. The key physics is illustrated in
\figref{fig:modes_o5} for the interaction between two perpendicularly
polarized, counterpropagating \Alfven waves with wavevectors
$\V{k}_1^+ = k_\perp \xhat - k_\parallel \zhat$ and $\V{k}_1^- =
k_\perp \yhat + k_\parallel \zhat$, denoted by the red circles in
panel (a).  In this section, $k_\perp$ and $k_\parallel$ are positive
constants, and the components of a particular wavevector are denoted
$k_x$, $k_y$, and $k_z$. The asymptotic solution is ordered by the
small parameter $\epsilon \sim |\V{z}^\pm|/v_A \ll 1$. The lowest
order nonlinear interaction creates an inherently nonlinear, purely
magnetic mode with wavevector $\V{k}_2^{(0)}= k_\perp \xhat + k_\perp
\yhat$ (green triangle). The nonlinear evolution is presented in panel
(b) of \figref{fig:modes_o5}, where Elsasser field amplitudes
$|\V{z}^\pm|$ are represented by Elssaser potentials, $\zeta^\pm$
\cite{Schekochihin:2009}, given by $\V{z}^{\pm} = \zhat \times
\nabla_\perp \zeta^{\pm}$.  The Elsasser potential is normalized by
$\zeta_{NL} \equiv k_\parallel v_A/k_\perp^2$ and time is normalized
by the primary \Alfven wave frequency $\omega_0=k_\parallel v_A$.  In
panel (b), the secondary mode (green) has $k_z=0$ and frequency
$\omega=2 \omega_0$, but does not grow secularly in time.  At next
order, the primary modes then interact with this secondary mode to
transfer energy secularly (with an amplitude that grows linearly in
time) to two nonlinearly generated \Alfven waves $\V{k}_3^+$ and
$\V{k}_3^-$ (blue squares), where $\V{k}_1^+ $ transfers energy to an
\Alfven wave with $\V{k}_3^+ = 2 k_\perp \xhat + k_\perp \yhat -
k_\parallel \zhat$, and $\V{k}_1^-$ transfers energy to $\V{k}_3^- =
k_\perp \xhat+ 2k_\perp \yhat + k_\parallel \zhat$.  This process is
the fundamental mechanism by which turbulence transfers energy from
large to small scales \cite{Howes:2013a}.

This fundamental energy transfer mechanism in the weakly nonlinear
limit, derived in the limit of incompressible MHD, has been confirmed
numerically with gyrokinetic numerical simulations in the MHD regime
\cite{Nielson:2013a} and verified experimentally in the laboratory
\cite{Howes:2012b,Howes:2013b,Drake:2013}, establishing \Alfven wave
collisions as the fundamental building block of astrophysical plasma
turbulence. This nonlinear mechanism is likely to be the physics
leading to a non-zero third-moment of turbulent magnetic field
measurements, a statistical measure often used to estimate the
turbulent energy cascade rate \cite{Coburn:2015}.

The validation of the incompressible MHD analytical solution for the
nonlinear evolution of \Alfven wave collisions with a kinetic
numerical simulation has two important implications.  First, the
properties of incompressible MHD solution persist even under the
weakly collisional plasma conditions relevant to realistic space and
astrophysical plasmas.  Second, the dynamical behavior of the
turbulent cascade in the MHD regime, $k_\perp \rho_i \ll 1$, in
particular the nonlinear energy transfer, is adequately described by
the simplified framework of incompressible MHD.  This second point
highlights that, even in a kinetic plasma, the nonlinear energy
transfer is essentially fluid in nature, and can be modeled as a
nonlinear wave-wave interaction. On the other hand, the physical
mechanisms responsible for damping of the turbulent fluctuations under
weakly collisional plasma conditions, as will be argued below, must be
essentially kinetic in nature, dominated by wave-particle
interactions.

I also note that the physics of \Alfven wave collisions highlights the
fact that plasma turbulence is inherently three-dimensional
\cite{Howes:2011a,Howes:2013a,Howes:2014c}. The linear term in
\eqref{eq:elsasserpm} that governs the propagation of the \Alfven
waves along the equilibrium magnetic field and is nonzero only when
the parallel wavenumber $k_\parallel \ne 0$, requiring the inclusion
of the field-parallel dimension. And the vector nature of the
nonlinear term requires variation in both of the directions
perpendicular to the magnetic field for the nonlinearity to be
nonzero.  The manifestly three-dimensional nature of plasma turbulence
not only applies to incompressible MHD plasmas, but persists as a
general characteristic of the turbulence for more complex plasmas,
such as compressible MHD plasmas or kinetic plasmas
\cite{Howes:2011a}.

%===================================================================
\subsection{What Causes the Anisotropic Cascade in Turbulent Plasmas?}

The anisotropic nature of the nonlinear energy transfer in \Alfvenic
plasma turbulence has long been recognized in laboratory plasmas
\cite{Robinson:1971,Zweben:1979,Montgomery:1981} and in the solar wind
\cite{Belcher:1971}, as well as in the results of numerical
simulations \cite{Shebalin:1983,Oughton:1994}. It is observed that
energy is preferentially transferred to small scales perpendicular to
the local magnetic field direction, leading to anisotropic,
small-scale turbulent fluctuations that are elongated along the local
magnetic field, characterized by the wavevector anisotropy
$k_\parallel \ll k_\perp$.

Why is magnetized plasma turbulence anisotropic? This question can be
answered rather simply, in the context of incompressible MHD, through
physical intuition derived from the picture of nonlinear energy
transfer presented above.  Basically, the anisotropy arises as a
direct consequence of two facts: (i) only counterpropagating \Alfven
waves interact nonlinearly, and (ii) the nonlinear energy transfer is
maximized for interactions between perpendicularly polarized \Alfven
waves.

Consider the nonlinear interaction between two plane \Alfven waves
with arbitrary wavevectors $\V{k}_1$ and $\V{k}_2$. We adopt the
convention that the wave frequency must be non-negative, $\omega\ge
0$, so that the propagation direction of the wave along the mean
magnetic field is given by the sign of the parallel component of the
wavevector.  The nonlinear interaction between these two modes is
nonzero only for counterpropagating \Alfven waves
\cite{Iroshnikov:1963,Kraichnan:1965,Howes:2013a}, so the parallel
components $k_{\parallel 1}$ and $k_{\parallel 2}$ must be opposite in
sign. A mode $\V{k}_3$ receiving energy via the nonlinear interaction
between these two modes must satisfy $\V{k}_3= \V{k}_1 + \V{k}_2$, so
its parallel component must have $k_{\parallel 3} \le k_{\parallel 1}$
and $k_{\parallel 3} \le k_{\parallel 2}$. The nonlinear term in
\eqref{eq:elsasserpm} has the form $\V{z}^{\mp}\cdot \nabla
\V{z}^{\pm}$. The magnitude of this nonlinear term is maximized when
the perpendicular components of the wavevectors of the interacting
\Alfven waves are orthogonal, $\V{k}_{\perp 1} \cdot \V{k}_{\perp
  2}=0$, so we expect the nonlinear energy transfer to be dominated by
interactions in which the colliding \Alfven waves are perpendicularly
polarized \cite{Howes:2013a}. For this case of orthogonal
perpendicular wavevector components, $\V{k}_{\perp 1} \cdot
\V{k}_{\perp 2}=0$, the resulting magnitude of the perpendicular
component of $\V{k}_3$ will satisfy $k_{\perp 3} \ge k_{\perp 1}$ and
$k_{\perp 3} \ge k_{\perp 2}$. In summary, the parallel wavevector
component of \Alfvenic fluctuations decreases or remains constant due
to the fact that only counterpropagating \Alfven waves interact
nonlinearly, while the perpendicular wavevector component of the
fluctuations generally increases due to the dominance of nonlinear
interactions between perpendicularly polarized \Alfven waves. Thus,
the resulting nonlinear energy transfer is expected to be anisotropic,
with energy preferentially flowing towards smaller scales in the
perpendicular direction than in the parallel direction.

The argument presented above is similar to a previous explanation of
anisotropic energy transfer
\cite{Shebalin:1983,Oughton:1994,Sridhar:1994,Montgomery:1995,Ng:1996,Galtier:2000}
which relies on the wavevector $\V{k}_3= \V{k}_1 + \V{k}_2$ and
frequency $\omega_3= \omega_1 + \omega_2$ matching conditions in the
weak turbulence limit for resonant interactions between \Alfven waves
with wavevectors $\V{k}_1$ and $\V{k}_2$. Retaining the convention of
non-negative frequencies, the linear \Alfven wave frequency is given
by $\omega = |k_\parallel |v_A$.  Therefore, the parallel component of
the wavevector matching condition and the frequency matching condition
lead to the equations $k_{\parallel 3}=k_{\parallel 1}+k_{\parallel
  2}$ and $|k_{\parallel 3}|=|k_{\parallel 1}|+|k_{\parallel
  2}|$. Since only counterpropagating \Alfven waves interact
nonlinearly, and thus $k_{\parallel 1}$ and $k_{\parallel 2}$ must be
opposite in sign (or zero), these equations have a solution only if
either $k_{\parallel 1}=0$ or $k_{\parallel 2}=0$.  This argument
establishes the necessity of modes with $k_\parallel=0$ in resonant
nonlinear interactions in the weak turbulence limit, but elucidates
neither the origin nor the nature of such modes. In summary, the parallel
wavenumber does not increase through nonlinear interactions in weak
incompressible MHD turbulence, while the perpendicular wavenumber
generally does increase, yielding an anisotropic transfer of energy to
modes with $k_\parallel \ll k_\perp$.

The analytical solution for \Alfven wave collisions in the weakly
nonlinear limit presented in \secref{sec:cascade} can be used to
illustrate this previous explanation for the anisotropic energy
transfer and to clarify the origin and role of the $k_\parallel=0$
modes.  Using the wavevector notation from \secref{sec:cascade}, where
$k_z$ represents the wavevector component parallel to the local mean
magnetic field, the nonlinear interaction between counterpropagating
plane \Alfven waves given by wavevectors $\V{k}_1^+$ and $\V{k}_1^-$
may lead to energy transfer to two other Fourier modes $\V{k}_2^{(0)}=
\V{k}_1^-+ \V{k}_1^+$ and $\V{k}_2^{(+2)}= \V{k}_1^-- \V{k}_1^+$.  The
detailed analytical solution \cite{Howes:2013a} demonstrates that the
resonant interactions that lead to the secular transfer of energy to
smaller scales are mediated by a nonlinearly generated mode with
$k_z=0$, so this result singles out $\V{k}_2^{(0)}$ as the key mode
mediating the nonlinear energy transfer.  For this particular problem,
the primary counterpropagating \Alfven waves $\V{k}_1^+$ and
$\V{k}_1^-$ each then interact nonlinearly with this self-consistently
generated mode $\V{k}_2^{(0)}$ to transfer energy resonantly to
\Alfven waves with wavevectors $\V{k}_3^+$ and $\V{k}_3^- $.  It is
these two subsequent interactions that correspond directly to the
previous explanation
\cite{Shebalin:1983,Oughton:1994,Sridhar:1994,Montgomery:1995,Ng:1996,Galtier:2000}
of anisotropic energy transfer in the weak turbulence limit. In the
presence of pre-existing energy in $k_z=0$ modes, this energy transfer
corresponds to a dominant three-wave interaction.  In the absence of
such pre-existing $k_z=0$ modes, it is the self-consistently generated
$\V{k}_2^{(0)}$ mode that mediates the energy transfer.  In this case,
there is no energy transfer via three-wave interactions, but rather
the combined interactions $\V{k}_2^{(0)}= \V{k}_1^-+ \V{k}_1^+$ and
$\V{k}_3^\pm = \V{k}_1^\pm + \V{k}_2^{(0)}$, together equivalent to a
four-wave interaction, constitute the dominant mechanism of nonlinear
energy transfer. Note that the mode $\V{k}_2^{(0)}$ has no associated
velocity fluctuation, $\V{u}=0$, but does include a non-zero
perpendicular magnetic field fluctuation, $\delta \V{B}_\perp \ne 0$
\cite{Howes:2013a}, so this mode has both $\V{z}^+\ne 0$ and
$\V{z}^-\ne 0$, and therefore interacts with \Alfven waves propagating
in either direction along the magnetic field. Physically, the mode
$\V{k}_2^{(0)}$ can be interpreted as a shear in the magnetic field
that the \Alfven waves traveling in either direction attempt to follow.

These arguments demonstrate that, in the weak turbulence limit, $\chi
\ll 1$, no parallel cascade occurs, yielding a maximally anisotropic
energy transfer only to smaller perpendicular scales.  Does the
physical picture of energy transfer mediated by self-consistently
generated $k_z=0$ modes persist in the strong turbulence limit, $\chi
\sim 1$? Let us consider an \Alfven wave collision problem equivalent
to the weakly nonlinear case presented in \secref{sec:cascade}, but
with the initial \Alfven wave amplitudes increased to yield a strongly
nonlinear case with $\chi=1$. Simulations of \Alfven wave collisions
in this strong turbulence limit \cite{Howes:2014d} demonstrate that
the energy transfer in perpendicular Fourier space flows primarily to
modes along the three diagonal lines represented in panel (a) of
\figref{fig:modes_o5}. Due to the wavevector matching conditions for
this particular problem, the parallel wavenumber\footnote{Recall that,
  in this notation, $k_\parallel$ is a positive constant describing
  the initial counterpropagating \Alfven waves.} is constant along
these diagonal lines, with values $k_z=+k_\parallel$, $k_z=0$, and
$k_z=-k_\parallel$, while the perpendicular wavenumber increases with
the distance from the origin of the plot. Thus, the energy transfer is
anisotropic in this strongly nonlinear case, consistent with previous
findings of anisotropic energy transfer in strong plasma turbulence.
Of course, in contrast to the weakly nonlinear case where the energy
transfer is dominated by interactions involving just the modes
$\V{k}_1^\pm$, $\V{k}_2^{(0)}$, and $\V{k}_3^\pm$, the strongly
nonlinear limit involves significant contributions from higher-order
terms in the asymptotic expansion. These terms lead to significant
nonlinear energy transfer involving many other three-wave couplings,
ultimately leading to a parallel transfer of energy to modes with
$|k_z| > k_\parallel$, but at a rate slower than the cascade of energy
to higher perpendicular wavenumber. The observed channeling of energy
along the three diagonal lines in panel (a) of \figref{fig:modes_o5}
suggests that $k_z=0$ modes continue to play a key role in mediating
the nonlinear energy transfer in strong turbulence. It is important to
note that, for an \Alfven wave collision with $\chi \sim 1$, the
self-consistently generated $k_z=0$ modes rise to amplitudes with the
same order of magnitude as the primary \Alfven waves, so pre-existing
energy in $k_z=0$ modes is unnecessary to yield a strong cascade of
turbulent energy.

In summary, the anisotropic cascade of energy in incompressible MHD
turbulence is a consequence of the two facts that only
counterpropagating \Alfven waves interact nonlinearly and that the
nonlinear term is greatest for interactions between perpendicularly
polarized \Alfven waves. Note this explanation of the anisotropy in
terms of the linear and nonlinear properties of the turbulent
fluctuations differs substantially from a statistical argument in this
issue \cite{Oughton:2015} relying on the coupling of third-order to
second-order correlations in the turbulence.  It is an open question
why the property of anisotropic energy transfer appears to persist
under less restrictive plasma conditions than incompressible MHD (in
particular for the dispersive modes at scales below the ion Larmor
radius) which allow for copropagating modes to interact nonlinearly.
An analytical solution of the nonlinear energy transfer in the
dispersive regime of kinetic \Alfven waves may help to answer this
open question.

%===================================================================
\subsection{What is Low-Frequency Turbulence in a Weakly Collisional Plasma?}

As a relevant aside, it is worthwhile to consider the nature of
low-frequency\footnote{Low frequency here refers to frequencies much
  smaller than the electron plasma frequency, $\omega \ll
  \omega_{pe}$.} turbulence in a weakly collisional plasma. Turbulence
arises as a result of nonlinearities in the equations of evolution for
the plasma.  The Boltzmann equation describes the evolution of the
six-dimensional distribution function $f_s(\V{r},\V{v},t)$ for a
plasma species $s$,
\begin{equation}
\frac{\partial f_s}{\partial t} + \mathbf{v}\cdot \nabla f_s + \frac
     {q_s}{m_s}\left[ \mathbf{E}+ \frac{\mathbf{v} \times \mathbf{B}
       }{c} \right] \cdot \frac{\partial f_s}{\partial \mathbf{v}} =
     \left(\frac{\partial f_s}{\partial t} \right)_{\mbox{coll}}.
\label{eq:boltzmann}
\end{equation}
A Boltzmann equation for each plasma species together with Maxwell's
equations form the closed set of Maxwell-Boltzmann equations governing
the dynamics of a kinetic plasma. Since Maxwell's equations are
linear, the only nonlinearities appearing in the Maxwell-Boltzmann
equations occur in the electromagnetic Lorentz force term and the
collisional term in \eqref{eq:boltzmann}. Under the weakly collisional
conditions relevant to the solar wind, the collisional term is
subdominant, and thus the physics of solar wind turbulence is
controlled by the electromagnetic Lorentz force on the particles.

The charge density and current density give rise to the
electromagnetic fields through Maxwell's equations, and therefore the
fields are entirely determined by the two lowest-order moments of the
ion and electron distribution functions. Thus, the forces in weakly
collisional plasma turbulence are essentially fluid in nature, since
they depend on only the low-order moments. For the non-relativistic
conditions of solar wind turbulence, the plasma fluctuations are
quasineutral, with typical charge density fluctuations, $\sum_s q_s
\delta n_s$, normalized by the total ion charge density, $q_i n_{0i}$,
having an order of magnitude $\sum_s q_s \delta n_s/q_i n_{0i} \sim
\mathcal{O} (v_A^2/c^2) \ll 1$ \cite{Howes:2006}. Therefore, with
negligible charge density fluctuations, the electromagnetic fields
mediating the low-frequency turbulence in the solar wind are
determined entirely by the current density, given by the first moments
of the ion and electron distribution functions.

The energy  in a volume of plasma is given by 
\begin{equation}
W = \int d^3\V{r} \left( \frac{|\V{E}|^2 + |\V{B}|^2}{8 \pi} + \sum_s
\int d^3\V{v}\frac{1}{2} m v^2 f_s \right),
\label{eq:vlasov_energy}
\end{equation}
the sum of the electromagnetic field energy and kinetic energy of the
plasma particles \cite{Krall:1973}.  Note that the kinetic energy
density of a plasma species is given by $\varepsilon_s = \int
d^3\V{v}\frac{1}{2} m v^2 f_s $, and we may define a ``kinetic''
temperature $T_{k s}$ (expressed in units of energy) for a monatomic
plasma species $s$ in three spatial dimensions by
\begin{equation}
 T_{k s} \equiv \frac{2}{3 n_s} \left( \varepsilon_s - \frac{1}{2} n_s m_s |\V{U}_s|^2\right), 
\end{equation}
where $\V{U}_s$ is the bulk fluid velocity of species $s$, given by
the first moment of the distribution function. Therefore, the energy in a 
kinetic plasma may be expressed by 
\begin{equation}
W = \int d^3\V{r} \left[ \frac{|\V{E}|^2 + |\V{B}|^2}{8 \pi} + \sum_s\left(
\frac{1}{2} n_s m_s |\V{U}_s|^2 +\frac{3}{2}n_s T_{k s}  
 \right) \right].
\label{eq:vlasov_energy2}
\end{equation}
Note that, if the distribution function for species $s$ is Maxwellian
(corresponding to local thermodynamic equilibrium), then the kinetic
temperature is equal to the thermodynamic temperature of that species,
$T_{k s}=T_s$. In this case, the last term in
\eqref{eq:vlasov_energy2} corresponds entirely to the thermal energy.
Energy cannot be extracted from a thermal distribution without
lowering the entropy, so this energy is thermodynamically
inaccessible. But, if the distribution function deviates from a
Maxwellian, as will generally be the case for a turbulent weakly
collisional plasma, then the last term contains not only thermal
energy but also non-thermal free energy associated with the deviations
from a Maxwellian distribution. These deviations generally have the
form of small-scale structure in velocity space. In summary, the first
term in \eqref{eq:vlasov_energy2} represents the electromagnetic
fluctuation energy, the second term represents the kinetic energy of
the bulk fluid velocity, and the third term includes both non-thermal
free energy in the distribution function as well as thermal energy.

The energy given in \eqref{eq:vlasov_energy2} is conserved by the
Maxwell-Boltzmann equations independent of collisionality. It is
worthwhile considering the various physical mechanisms that can lead
to the transfer of energy among the different terms in
\eqref{eq:vlasov_energy2}. The low-frequency linear wave response of
the kinetic plasma leads to the continual exchange of energy between
the electromagnetic energy and bulk kinetic energy---the magnetic
tension in \Alfven waves, for example, leads to the periodic transfer
of energy between the magnetic energy and the kinetic energy of the
bulk plasma motion, an exchange that is purely fluid in
nature. Collisionless wave-particle interactions, such as Landau
damping, lead to the transfer of energy from the electromagnetic
fields into non-thermal free energy in the distribution
function. Conversely, unstable distribution functions can lead to the
transfer of energy from the non-thermal free energy in the velocity
distributions to electromagnetic fluctuations via collisionless
wave-particle interactions, such as kinetic temperature anisotropy
instabilities \cite{Gary:2015}. Both of these physical mechanisms of
energy transfer by wave-particle interactions are essentially kinetic
in nature. In addition to conserving energy, all of the processes
mentioned above also conserve entropy since the collisionless
Boltzmann equation (or the Vlasov equation) conserves
entropy. Collisions are the final physical mechanism that converts
non-thermal free energy in the distribution function to thermal
energy, a process that increases the entropy of the plasma, thereby
realizing irreversible thermodynamic heating of the
plasma. Collisions, of course, do not transfer energy between terms in
\eqref{eq:vlasov_energy2}, but merely convert non-thermal to thermal
energy within the last term in the equation.

Based on these considerations, here I suggest a practical definition
for what constitutes the turbulence in a kinetic plasma. The
turbulence is represented by the first moments of the distribution
functions, specifically the electromagnetic fluctuations (determined
by the current density) and the bulk fluid velocities of the plasma
species, corresponding to the first two terms in
\eqref{eq:vlasov_energy2}. The nonlinear interactions underlying the
turbulent cascade of energy from large to small scales are mediated by
the electromagnetic fields, and since these fields depend only on the
first moments of the ion and electron distribution functions, the
nonlinear dynamics of the turbulence is therefore fluid in nature.
Particle motions contributing to the third term, containing both
non-thermal and thermal energy in the distribution function, are not
part of the turbulence by this definition.  Collisionless
wave-particle interactions transfer energy from the turbulent
electromagnetic and bulk velocity fluctuations (the first two terms in
\eqref{eq:vlasov_energy2}) into non-thermal free energy in the plasma
species distribution functions (the third term).  This process damps
the turbulence, increasing the ``kinetic'' temperature, but does not
lead to thermodynamic heating of the plasma (an increase of the
thermal temperature). The physical mechanisms that damp the turbulent
fluctuations are therefore kinetic in nature. Collisions then act to
smooth out the non-thermal fluctuations in the velocity distribution,
irreversibly converting the non-thermal free energy in the
distribution function to thermal energy. Thus, the dissipation of
turbulence in a weakly collisional plasma is necessarily a two-step
process, whereby first entropy-conserving collisionless wave-particle
interactions damp the turbulent electromagnetic and bulk velocity
fluctuations, and then entropy-increasing collisions thermalize the
energy removed from the turbulent fluctuations. Other forms of particle
energization, such as the development of beams or high-energy tails in
the velocity distribution, may also occur via collisionless
wave-particle interactions, and correspond to the transfer of energy
from the electromagnetic fields to non-thermal free energy in the
distribution functions, a transfer that is also essentially kinetic in
nature.

%===================================================================
\subsection{Why are Parametric Instabilities Subdominant?}

The inherently anisotropic nature of plasma turbulence, with turbulent
fluctuations that satisfy $k_\parallel \ll k_\perp$, also leads to the
prediction that the nonlinearity associated with \Alfven wave
collisions dominates over other potential nonlinear mechanisms driven
by gradients parallel to the magnetic field, in particular parametric
instabilities, such as the
decay,\cite{Galeev:1963,Sagdeev:1969,Hasegawa:1976a,
  Derby:1978,Goldstein:1978,Spangler:1982,Sakai:1983,Spangler:1986,
  Terasawa:1986,Jayanti:1993,Hollweg:1994,Shevchenko:2003,Voitenko:2005}
modulational,\cite{Lashmore-Davies:1976,Mjolhus:1976,Mio:1976b,
  Sakai:1983,Wong:1986,Hollweg:1994,Shukla:2007} and
beat\cite{Wong:1986,Hollweg:1994} instabilities.  For example, even
for oblique \Alfven wave modes, numerical studies using both MHD
\cite{DelZanna:2001b,DelZanna:2014} and kinetic ion/fluid electron
hybrid \cite{Matteini:2010} simulations find parametric instability
growth rates that are proportional to $k_\parallel= k \cos\theta$,
where $\theta$ is the angle between the wavevector and the magnetic
field.  The root of this property is that the nonlinearity associated
with parametric instabilities is proportional to $ k_\parallel \delta
v$, whereas the nonlinearity associated with \Alfven wave collisions
is proportional to $ k_\perp \delta v$. Therefore, the relative
magnitude of the effect of parametric instabilities to that of \Alfven
wave collisions is $k_\parallel/k_\perp \ll 1$, so parametric
instabilities are expected to be subdominant in anisotropic plasma
turbulence \cite{Howes:2013a}. A more thorough investigation is
warranted to confirm this prediction.

%===================================================================
\subsection{Why is Critical Balance Natural for Turbulent Plasmas?}
The nonlinearity parameter, $\chi \equiv |\V{z}^{-}\cdot \nabla
\V{z}^{+}|/|\V{v}_A \cdot \nabla \V{z}^{+}|$, measures the strength of
incompressible MHD turbulence using the ratio of the magnitude of the
nonlinear term to that of the linear term in \eqref{eq:elsasserpm}.
In the weak turbulence limit, $\chi \ll 1$, the dynamics may be
rigorously calculated analytically using perturbation theory
\cite{Ng:1996,Galtier:2000,Lithwick:2003,Howes:2013a}, with the
results of such a calculation described above in
\secref{sec:cascade}. Strong incompressible MHD turbulence is
conjectured to maintain a state of critical balance
\cite{Goldreich:1995} in which the nonlinear and linear terms are of
equal magnitude, or $\chi \sim 1$. Although the concept of critical
balance is not universally accepted, a significant number of studies
employing various techniques to estimate the variation along the
\emph{local} mean magnetic field direction support critical balance
\cite{Cho:2000,Maron:2001,Horbury:2008,Podesta:2009a,Wicks:2010,Forman:2011,TenBarge:2012a,TenBarge:2013b},
while contradictory studies have uniformly used a global mean field
\cite{Matthaeus:1998,Tessein:2009,Grappin:2010}, an approach that has
been shown to give misleading results \cite{Cho:2000}.  Since the
linear terms are equal in magnitude to the nonlinear terms in
critically balanced strong turbulence, the linear plasma physics
continues to have an important impact on the dynamics in magnetized
plasma turbulence, unlike hydrodynamic turbulence. An important
consequence of this fact is that the turbulent fluctuations may
exhibit some linear mode properties, properties which may be exploited
to illuminate the nature of plasma turbulence
\cite{Klein:2012,Howes:2014b}. The relevance of linear physics to
plasma turbulence is responsible for the numerical finding that the
picture of energy transfer to smaller scales presented in
\secref{sec:cascade}, derived rigorously in the weakly nonlinear limit
\cite{Howes:2013a}, remains qualitatively correct, even in the limit
of strong, critically balanced turbulence, well beyond the formal
regime of applicability of the solution \cite{Howes:2014d}.

It is worthwhile to explain physically why critical
balance---equivalently described by a state in which the
characteristic linear timescale $\tau \sim l_\parallel/ v_A$ and
nonlinear timescale $\tau_{nl} \sim l_\perp /\delta v_\perp$ are in
balance---naturally arises through the dynamics of a turbulent plasma.
In these variables, $\chi = ( l_\parallel \delta v_\perp)/( l_\perp
v_A)$.  Consider an MHD plasma, initially at rest, that is shaken
transverse to the magnetic field at one position with a sinusoidal
variation of velocity of amplitude $\delta v_\perp$ over a distance
$l_\perp$. The timescale associated with this perturbation of the
plasma (and of the frozen-in magnetic field) is $\tau\sim l_\perp
/\delta v_\perp$.  Magnetic tension will lead to the propagation of
\Alfven waves up and down the magnetic field, driven by the applied
perturbation.  One may ask, for the period of forcing $\tau$, what is
the parallel wavelength of the resulting waves?  The parallel
wavelength will be given by $l_\parallel=v_A\tau = v_A ( l_\perp
/\delta v_\perp)$, so this leads to the relation $l_\parallel/ v_A =
l_\perp /\delta v_\perp$, or $\chi=1$.  If the transverse perturbation of the
plasma is due to the nonlinear terms in \eqref{eq:elsasserpm} such
that $\tau_{nl} \sim l_\perp /\delta v_\perp$, then this leads to the
relation between the linear and nonlinear timescales, $\tau \sim
\tau_{nl}$.  Alternatively, critical balance can be expressed as a
balance between frequencies $\omega \sim \omega_{nl}$, or $k_\parallel
v_A \sim k_\perp \delta v_\perp$.

One may wonder, why is the limit $\chi \gg 1$ not considered?  In this
case, the nonlinear term is much larger than the linear term, which
occurs when $l_\parallel \rightarrow \infty$, or $k_\parallel
\rightarrow 0$, the two-dimensional limit where fluctuations
perpendicular to the magnetic field are correlated for long distances
along the magnetic field.  But, transverse fluctuations in a turbulent
plasma will only remain correlated along a magnetic field if they an
exchange information along the field.  Since information propagates
along the magnetic field at the \Alfven speed, for a transverse
oscillation with a period $\tau\sim l_\perp /\delta v_\perp$, the
distance along the magnetic field over which the transverse motions
can remain correlated must have $l_\parallel \le v_A \tau = v_A (
l_\perp /\delta v_\perp)$.  Therefore, one obtains the constraint
$l_\parallel/ v_A \le l_\perp /\delta v_\perp$, or in terms of the
nonlinearity parameter, $\chi \le 1$.  Therefore, fluctuations with
$\chi \gg 1$ will decorrelate and evolve toward a state with $\chi \le
1$. Thus, the turbulence cannot be any more two dimensional than
allowed by the critical balance because fluctuations in any two planes
perpendicular to the mean field can only remain correlated if an
\Alfven wave can propagate between them in less than their
perpendicular decorrelation time \cite{Maron:2001,Schekochihin:2009}.

In closing, it is worthwhile pointing out the common misconception
that plasma turbulence in the solar wind must be weak because the
turbulent magnetic field fluctuations $\delta B$ are small compared to
the mean magnetic field magnitude $B_0$, or $\delta B/B_0 \ll 1$. In
fact, the strength of the turbulence, estimated by the nonlinearity
parameter $\chi \sim k_\perp \delta B_\perp/ (k_\parallel B_0)$ as the
ratio of the nonlinear to the linear term in \eqref{eq:elsasserpm},
depends not only on the normalized amplitude of the fluctuations,
$\delta B/B_0$, but also on the anisotropy of the fluctuations,
$k_\perp /k_\parallel$. Direct multi-spacecraft measurements of
turbulent magnetic field fluctuations at the small-scale end of the
inertial range find $k_\perp /k_\parallel \gtrsim 10$
\cite{Sahraoui:2010b,Narita:2011,Roberts:2013}, so even the small
amplitude magnetic field fluctuations measured at that scale,
typically of magnitude $\delta B/B_0\sim 0.1$, lead to strong
nonlinear interactions.

%===================================================================
\subsection{What is the Nature of the Compressible Turbulent Fluctuations?}
In addition to the incompressible dynamics of \Alfven waves that
dominate turbulence in the inertial range of magnetized plasmas,
plasma turbulence also contains a small fraction (typically 10\% or
less) of energy in compressible fluctuations \cite{Bruno:2005}.
Recent high-frequency measurements of the density fluctuations are
presented elsewhere in this issue \cite{Riazantseva:2015}.  Exploiting
the properties of compressible fluctuations in the inertial range
using linear kinetic theory, the nature of the compressible
fluctuations may be identified by analyzing the correlation between
the density and parallel magnetic field fluctuations, \XCC.  A recent
comparison of \XCC between 10 years of WIND spacecraft data at 1~AU
and theoretical predictions using the synthetic spacecraft data method
\cite{Klein:2012} has shown that there is negligible energy in the
kinetic counterpart of the fast magnetosonic mode, and that the
observed spectrum of compressible energy is composed entirely of
kinetic slow mode\footnote{Future work to confirm this finding should
  address the issue of whether modes driven by the kinetic mirror
  instability are also consistent with the WIND observations, but the
  ubiquity of the compressible fluctuations measured suggest that they
  are not restricted to plasma intervals with the temperature
  anisotropy $T_{\perp p}/T_{\parallel p} > 1$ necessary for the
  mirror instability to generate fluctuations.}  fluctuations
\cite{Howes:2012a}.  This lack of energy in kinetic fast waves remains
unexplained, but may be due to shock dissipation of fast wave energy
in the inner heliosphere or to little generation of fast wave energy
in the coronal plasma that is launched out into the solar wind.  In
the limit of significant anisotropy $k_\parallel \ll k_\perp$, it has
been shown theoretically
\cite{Schekochihin:2009,Howes:2013a,Howes:2014c} and numerically
\cite{Maron:2001} that the turbulent slow mode fluctuations (or
pseudo-\Alfven waves in the case of incompressible MHD) are cascaded
passively by the \Alfvenic fluctuations, but that the slow waves do
not impact the cascade of \Alfven waves. Therefore, the turbulent
fluctuations in the solar wind inertial range consist of an active
cascade of incompressible, counterpropagating \Alfven waves and a
passive cascade of compressible slow wave fluctuations.

%===================================================================
\subsection{How Does Temperature Anisotropy Arise from or Affect Turbulence?}

Recent numerical \cite{Servidio:2014a} and theoretical
\cite{TenBarge:2014c} results suggest that the compressible
fluctuations in turbulence may control the observed spread of proton
temperature anisotropy $T_{\perp i}/T_{\parallel i}$ in the solar wind
plasma.  The observations of $T_{\perp i}/T_{\parallel i}$ in the
solar wind are distributed across the broad range of values between
the kinetic instability boundaries due to the mirror instability at
$T_{\perp i}/T_{\parallel i}>1$ and the oblique firehose instability
at $T_{\perp i}/T_{\parallel i}<1$
\cite{Gary:1994,Gary:2001,Kasper:2002,Marsch:2004,Hellinger:2006,Matteini:2007,Bale:2009}.
Although a wide range of physical mechanisms affect $T_{\perp
  i}/T_{\parallel i}$---including collisions, ion parallel heat flux,
spherical expansion, radial compression and expansion, turbulent
heating, turbulent compression, and kinetic
instabilities---theoretical considerations show that kinetic slow mode
fluctuations may be responsible for the observed spread of $T_{\perp
  i}/T_{\parallel i}$ values \cite{TenBarge:2014c}, and recent kinetic
numerical simulations are consistent with this explanation
\cite{Servidio:2014a}.

In addition to the proton temperature anisotropy being affected by the
turbulence, when $T_{\perp i}/T_{\parallel i}$ exceeds certain
threshold values, kinetic temperature anisotropy instabilities can
generate fluctuations\footnote{In addition to temperature anisotropy
  instabilities, other kinetic instabilities, such as those driven by
  beams or the relative drift between different plasma species, can
  also give rise to electromagnetic fluctuations.} at characteristic
ion kinetic length scales, thereby injecting energy into the turbulent
plasma, as reviewed elsewhere in this issue \cite{Gary:2015}.  Here I
suggest that the fluctuations driven by such kinetic instabilities are
distinct from the main body of turbulent fluctuations associated with
the cascade of energy from large to small scales.  The distinction is
that the instability-driven fluctuations inhabit a completely
different region of wavevector space than the fluctuations arising
from the turbulent cascade (shown in \figref{fig:model}), although to
distinguish them from single-point spacecraft measurements requires
careful analysis.

There exist four ion temperature anisotropy instabilities that may
serve to inject energy into the turbulent plasma: the parallel (or
whistler) firehose instability \cite{Kennel:1966,Gary:1976}, the
\Alfven (or oblique) firehose instability \cite{Hellinger:2000}, the
mirror instability \cite{Vedenov:1958,Tajiri:1967,Southwood:1993}, and the proton
cyclotron instability \cite{Gary:1976}. For each of these
instabilities, unstable wave growth peaks at $k_\parallel d_i \sim 1$,
where $d_i=c/\omega_{pi} = v_A/\Omega_i$ is the ion inertial length.
For the parallel firehose and proton cyclotron instabilities, these
waves have $k_\perp d_i \ll 1$, and for the \Alfven firehose and
mirror instabilities, they have $k_\perp d_i \sim 1$.  Therefore,
these instabilities inhabit a combined region with $k_\parallel d_i
\sim 1$ and $k_\perp d_i \lesssim 1$, with a typical wavevector
anisotropy of $k_\parallel /k_\perp \gtrsim 1$. In contrast, at the
small scales around the ion inertial length, turbulent fluctuations in
a magnetic plasma have a typical wavevector anisotropy $k_\parallel/k_\perp \ll 1$
\cite{Goldreich:1995,Boldyrev:2006,Howes:2008b,Schekochihin:2009}.
Therefore, the instability-driven fluctuations do not occupy the same
region of wavevector space as the fluctuations of the turbulent
cascade from large scales. 

Note that kinetic instabilities of the electrons can also inject
fluctuation energy into the solar wind at yet smaller scales, and
recent work has indeed identified the signature of localized whistler
wavepackets at scales $k d_i \gg 1$ which appear to be driven by the
whistler heat flux instability \cite{Lacombe:2014}.  These whistler
modes have quasi-parallel wavevectors with $k_\parallel \gg k_\perp$,
and therefore also inhabit a region of wavevector space distinct from
the anisotropic cascade of energy from large scales.

Discriminating fluctuations with $k_\parallel /k_\perp \gtrsim 1$ from
those with $k_\parallel /k_\perp \ll 1$, however, is not generally
possible using single-point spacecraft measurements. Such measurements
only provide the projection of the wavevector along the solar wind
velocity, so there is insufficient information to separate the
parallel from the perpendicular component of the wavevector.  Only
when the magnetic field direction is parallel to the solar wind flow,
a relatively rare occurrence in the near-Earth solar wind, can one
uniquely determine the value of $k_\parallel$. Measurements of the
fluctuating magnetic helicity $\sigma_m$ as a function of the angle
$\theta$ between the solar wind velocity $\V{v}_{sw}$ and $\V{B_0}$
indeed have discovered two distinct signatures in regions
perpendicular and parallel to $\V{B_0}$ at frequencies $f \sim 1$ Hz
\cite{Podesta:2011a,He:2011}, corresponding to $k d_i \sim 1$. It has
been suggested that the perpendicular signature arises from kinetic
\Alfven waves with $k_\parallel \ll k_\perp$, while the parallel
signature arises from either ion cyclotron waves propagating away from
the sun or whistler waves propagating toward to the sun with
$k_\parallel \gg k_\perp$
\cite{He:2011,Podesta:2011a,Podesta:2011b}. An analysis comparing
predictions from the synthetic spacecraft data method
\cite{Klein:2012} to observations confirms that such a two-component
model---consisting primarily of kinetic \Alfven waves with
$k_\parallel \ll k_\perp$ with a small amount (5\% in energy) of
either ion cyclotron or whistler waves with $k_\parallel \gg
k_\perp$---indeed reproduces the observations, supporting the
hypothesis that kinetic-instability-driven waves persist in the solar
wind alongside the \Alfvenic turbulent fluctuations that mediate the
nonlinear transfer of energy from large to small scales
\cite{Klein:2014a}.

Are these instability-driven parallel waves turbulent themselves (do
they transfer energy nonlinearly to small scales?\cite{Gary:2015}), or
do they merely persist in the turbulent environment caused by the
anisotropic \Alfvenic cascade of energy from large to small scales?
That the parallel waves have small amplitude and are propagating
unidirectionally suggests they do not interact nonlinearly among
themselves \cite{Klein:2014a}.  In addition, since the \Alfvenic wave
frequencies are linearly proportional to the parallel wavenumber,
$\omega \propto k_\parallel$, the instability-driven waves with
$k_\parallel d_i \sim 1$ have much higher frequencies than the
anisotropic fluctuations of the turbulent cascade that have
$k_\parallel d_i \ll 1$. Here I suggest that this frequency mismatch
impedes significant nonlinear coupling of the parallel waves to the
turbulent fluctuations, so the parallel waves are not turbulent but
merely persist alongside the anisotropic \Alfvenic
turbulence. Therefore, the kinetic instabilities and their resulting
fluctuations, distinguished observationally by the parallel magnetic
helicity signature, operate separately from the turbulent cascade of
energy from large to small scales.

%===================================================================
\subsection{How Are Coherent Structures Generated in Plasma Turbulence?}
\label{sec:curr}
Numerical simulations of plasma turbulence demonstrate the ubiquitous
development of coherent structures, specifically current sheets, at
small scales
\cite{Biskamp:1989,Spangler:1998,Spangler:1999,Biskamp:2000,Maron:2001,Merrifield:2005},
and dissipation is largely concentrated in these current sheets
\cite{Wan:2012,Karimabadi:2013,TenBarge:2013a,Wu:2013a,Zhdankin:2013}.
There has been a flurry of activity recently seeking evidence for such
localized heating through statistical analyses of solar wind
observations
\cite{Osman:2011,Borovsky:2011,Osman:2012a,Osman:2012b,Perri:2012a,Wang:2013,Wu:2013a,Osman:2014b}. But
one fundamental question remains unanswered: what mechanism governs
the generation of current sheets in plasma turbulence?  New research
suggests that current sheets arise naturally as a consequence of
\Alfven wave collisions in the strong  turbulence limit
\cite{Howes:2014d}.

Consider first the physics of the \Alfven wave collision problem
described in \secref{sec:cascade} in the weak turbulence limit, $\chi
\ll 1$.  For initial \Alfven waves that have equal amplitudes given by
the constants $z_+=z_-$, the nonlinearity parameter is $\chi=k_\perp
z_\pm/(k_\parallel v_A)$. An asymptotic expansion of the nonlinear
evolution in the weakly nonlinearly limit can be performed, ordered by
the small expansion parameter $\epsilon = z_\pm/v_A \ll 1$
\cite{Howes:2013a}. In this limit, the physics of the nonlinear energy
transfer is dominated by just five modes as depicted in panel (a) of
\figref{fig:modes_o5}: the primary \Alfven waves $\V{k}_1^\pm$ at
order $\epsilon$ (red circles), the secondary mode $\V{k}_2^{(0)}$ at
order $\epsilon^2$ (green triangles), and the tertiary \Alfven waves
$\V{k}_3^\pm$ at order $\epsilon^3$ (blue squares). The secular
transfer of energy to small scales is accomplished by just three
nonlinear interactions, the interaction $\V{k}_2^{(0)}= \V{k}_1^-+
\V{k}_1^+$ at order $\epsilon^2$ and the two interactions $\V{k}_3^\pm
= \V{k}_1^\pm + \V{k}_2^{(0)}$ at order $\epsilon^3$.  These nonlinear
interactions, governed by the form of the nonlinear terms in
\eqref{eq:elsasserpm}, determine the phase and amplitude relationships
among these five relevant Fourier modes.  For example, for the
particular problem described in \secref{sec:cascade}, the nonlinearly
generated \Alfven waves $\V{k}_3^\pm$ are phase-shifted from the
primary \Alfven waves $\V{k}_1^\pm$ by $90^\circ$
\cite{Howes:2013a,Howes:2013b}. In fact, recent analytical and
numerical work argues for the importance of intermittency, or phase
synchronization, in weak incompressible MHD turbulence
\cite{Meyrand:2014}.

As the initial \Alfven wave amplitudes $z_\pm$ increase to the
strongly nonlinear regime, $\chi \rightarrow 1$, the asymptotic
expansion of the equations of evolution ceases to be well-ordered, and
higher-order terms---terms that can safely be neglected in the weakly
nonlinearly limit---begin to contribute significantly. The nonlinear
interactions associated with these higher-order terms mediate
significant energy transfer to additional Fourier modes in the
perpendicular plane, as depicted in \figref{fig:modes_o5}.  Yet, even
in the strong turbulence limit, the phase and amplitude relationships
among all of these modes remain completely specified by the nonlinear
terms in the equations of evolution.

Nonlinear gyrokinetic simulations using the Astrophysical Gyrokinetics
Code, \T{AstroGK} \cite{Numata:2010}, have demonstrated the
development of a current sheet as a natural consequence of the
nonlinear evolution of an \Alfven wave collision in the strong
turbulence limit, $\chi \sim 1$ \cite{Howes:2014d}. Through these
simulations, the qualitative understanding of the nonlinear energy
transfer gleaned from the analytical solution in the weakly nonlinear
limit $\chi \ll 1$ \cite{Howes:2013a} is found to persist as one
approaches and reaches the strongly nonlinear limit, $\chi \rightarrow
1$. In particular, the nonlinearly generated $k_z=0$ modes continue to
play a key role in the energy transfer. It is found that the energy
transfer to higher perpendicular wavenumbers flows predominantly along
the three diagonal lines in panel (a) of \figref{fig:modes_o5},
largely confined to these modes with $k_z=+k_\parallel$, $k_z=0$, and
$k_z=-k_\parallel$ \cite{Howes:2014d}. Generally, it appears that the
higher-order $k_z=0$ modes are also purely magnetic, inherently
nonlinear modes, similar to $\V{k}_2^{(0)}$.  And, the energy in the
higher-order modes with $k_z=\pm k_\parallel$ appears to primarily
consist of counterpropagating \Alfven waves, where the perpendicular
wavenumber of these \Alfven waves increases, given by the distance of
the mode from the origin of the plot.  Further analytical work is
necessary is necessary to confirm the nature of these higher-order,
nonlinearly generated modes. The modes along these three diagonals
represent a strikingly small fraction of the total possible
perpendicular Fourier modes, typically a few percent. It is
constructive interference among just this small number of modes that
results in the appearance of a coherent structure, in this case a
current sheet.

All of the modes shown in panel (a) of \figref{fig:modes_o5} can, in
principle, be computed analytically by extending the asymptotic
calculation to higher order, but the such a calculation involves a
rapidly growing number of terms. However, the confinement of the
turbulent energy to a few modes at each order $\epsilon^n$ can provide
a guide to limit the effort of detailed analytical calculations to
just the nonlinear modes directly involved in the dominant flow of
energy to small scales along the diagonal lines in
\figref{fig:modes_o5}. Another complication is that, as the
nonlinearity parameter increases to the strong turbulence limit, $\chi
\rightarrow 1$, the convergence of the asymptotic series becomes poor,
and the calculation must be taken to increasingly higher powers of
$\epsilon$.  Nevertheless, the onset of current sheet development in a
moderately nonlinear numerical calculation with $\chi=1/2$ is well
reproduced by an analytical calculation employing only the five modes
up to and including order $\epsilon^3$ \cite{Howes:2014d}.  This
result suggests that the physical mechanism of nonlinear energy
transfer in \Alfven wave collisions, outlined in \secref{sec:cascade},
is indeed ultimately responsible for the development of current sheets
in plasma turbulence, and that such coherent structure development can
be computed analytically.

Note that any coherent structure can be assembled through constructive
interference among spatial Fourier modes with particular amplitudes
and phases. For example, a square wave can be expressed as the sum of
sinusoidal modes $f(x) = \sin x + \sin(3x)/3 + \sin(5x)/5 + \ldots$.
In the \Alfven wave collision problem, it is constructive interference
among wave modes in the perpendicular plane---specifically the Fourier
modes depicted in \figref{fig:modes_o5}, where the amplitudes and
phases of these modes are determined by the form of the nonlinear
terms and can be computed analytically---that results in the
development of a coherent structure.  The observation of coherent
structures in plasma turbulence arising in the form of current sheets
is often invoked as evidence that their presence is inconsistent with
the strongly interacting \Alfven wave interpretation of plasma
turbulence. The results reviewed here suggest not only that \Alfven
waves are consistent with, but also that they are directly responsible
for, the development of current sheets. The development of a current
sheet is a natural consequence of the nonlinear interaction between
two counterpropagating \Alfven waves in the strong turbulence limit, a
result that is not \emph{a priori} obvious in any way.  This key
finding resolves the apparent contradiction between investigations
that view plasma turbulence as a sea of nonlinearly interacting
\Alfven waves and those that focus on the development of coherent
structures and their role as sites of enhanced dissipation.
Ultimately, this investigation of the nonlinear dynamics of \Alfven
waves enables a first-principles prediction of current sheet
development in plasma turbulence through the linear superposition
of the nonlinearly generated modes, and illuminates the nature of
these current sheets as constructive interference between
counterpropagating \Alfven waves.

Along with the picture of nonlinear energy transfer mediated by
counterpropagating \Alfven waves, this discovery of
Alfv\'en-wave-driven current sheet development completes the
foundation necessary to construct the first dynamical model of plasma
turbulence that self-consistently combines the physics of \Alfven
waves and small-scale current sheets, as presented below.  This
dynamical understanding of the origin and nature of current sheets
provides important constraints on how such coherent structures may be
dissipated, to be discussed in next section.

Before ending our discussion of current sheets, it is important to
point out that there are other mechanisms, such as particular flow and
magnetic field geometries, that can produce current sheets.  The
Orzsag-Tang vortex is one such example \cite{Orszag:1979}. Do such
alternative mechanisms play any role in the development of current
sheets in solar wind turbulence, or are strong \Alfven wave collisions
sufficient to account for all of the current sheets observed in plasma
turbulence?  To answer this open question, the properties of current
sheets generated by \Alfven wave collisions must be carefully compared
to those arising in numerical simulations and inferred from solar wind
observations.

%===================================================================
\section{The Damping of Turbulent Fluctuations and Plasma Heating}
\label{sec:diss}
At the forefront of plasma turbulence research stands the effort to
identify and characterize the physical mechanisms responsible for the
dissipation of the turbulent fluctuations and the conversion of their
energy to heat, or other energization, of the protons, electrons, and
minor ions. As discussed above, for a weakly collisional plasma, the
dissipation of turbulence necessarily includes two stages: first, the
entropy-conserving, collisionless damping of the turbulent
fluctuations; then, the ultimate conversion of the damped non-thermal
free energy through entropy-increasing collisions into plasma heat.

%Since the Boltzmann $H$ theorem demands
%that the increase of entropy in a weakly collisional plasma
%necessarily requires collisions \cite{Howes:2006}, the question of the
%physical mechanism of turbulent dissipation really boils down to the
%mechanism responsible for the collisionless damping of the turbulent
%fluctuations.

There are presently three leading candidates for the damping of the
turbulent fluctuations: (i) coherent wave-particle interactions, in
particular Landau damping\footnote{I include in my use of the term
  ``Landau damping'' all damping associated with the Landau resonance,
  thereby also including transit-time damping, the magnetic analogue
  of Landau damping.}  \cite{Leamon:1998b,Quataert:1998,Leamon:1999,
  Quataert:1999,Leamon:2000,Howes:2008b,Schekochihin:2009,TenBarge:2013a};
(ii) stochastic wave-particle interactions
\cite{Johnson:2001,Chen:2001,
  White:2002,Voitenko:2004,Bourouaine:2008,Chandran:2010a,Chandran:2010b,Chandran:2011,Bourouaine:2013};
and (iii) dissipation associated with the current sheets that are
ubiquitously observed in plasma turbulence
\cite{Dmitruk:2004,Markovskii:2011,Matthaeus:2011,Osman:2011,Servidio:2011a,Osman:2012a,Osman:2012b,Wan:2012,Karimabadi:2013,Osman:2014a,Osman:2014b}.
Observational constraints on the physical mechanisms responsible for
the dissipation of the turbulence and resulting plasma heating are
reviewed elsewhere in this issue \cite{Goldstein:2015}.  Note that
different mechanisms may dominate from one environment to another
(corona, inner heliosphere, outer heliosphere) due to variations in
the plasma parameters ($\beta$, $T_i/T_e$, etc.)  or changing
characteristics of the turbulence (scale and amplitude of the energy
injection, size of inertial range, etc.).  Below I present a dynamical
model for the cascade of anisotropic \Alfvenic turbulence down to and
below ion kinetic length scales, its collisionless damping by coherent
wave-particle interactions, and the ultimate conversion of the
turbulent energy into plasma heat.  In addition, I discuss how
stochastic heating and dissipation in current sheets relate to this
model.

%===================================================================
\subsection{A Model of Anisotropic \Alfvenic Turbulence}
\label{sec:alfturb}
The observed one-dimensional magnetic energy frequency spectrum in the
solar wind demonstrates a power law with spectral index $-5/3$ at low
frequencies \cite{Goldstein:1995}, a break at around $f \sim 0.4$~Hz,
and a steeper spectrum at higher frequencies with a spectral index of
approximately $-2.8$
\cite{Sahraoui:2009,Kiyani:2009,Alexandrova:2009,Chen:2010b,Sahraoui:2010b}.
The low-frequency range is denoted the \emph{inertial range} of solar
wind turbulence, the break is believed to be associated with an ion
kinetic length scale, and the high-frequency range is commonly denoted
the \emph{dissipation range}, although whether dissipation is actively
occurring at all scales in this range remains an open question. In
addition, multi-spacecraft observations show that the turbulent
fluctuations at the small scales near the break are highly
anisotropic, with $k_\parallel \ll k_\perp$
\cite{Sahraoui:2010b,Narita:2011,Roberts:2013}.

The following model of critically balanced, strong \Alfvenic
turbulence has been proposed to explain these observed features and to
provide a foundation upon which predictive models of solar wind
turbulence, its dissipation, and the resulting plasma heating may be
constructed
\cite{Howes:2008b,Howes:2008c,Schekochihin:2009,Howes:2011b,Howes:2013a,Howes:2014d}.
In this model, the inertial range occurs at scales $k_\perp \rho_i \ll
1$ and consists of counterpropagating \Alfven waves that transfer
energy to successively smaller scales via strong \Alfven wave
collisions.  At the scale $k_\perp \rho_i \sim 1$, the linear wave
physics transitions from that of nondispersive MHD \Alfven waves to
dispersive kinetic \Alfven waves, leading to the steepening of the
energy spectrum. At smaller scales, I suggest here that the nonlinear
energy transfer continues to be mediated by counterpropagating wave
collisions, only between kinetic \Alfven waves rather than MHD \Alfven
waves.\footnote{The growing body of observational evidence in support
  of kinetic \Alfven waves as the dominant contributor to the
  dissipation range of solar wind turbulence has recently been
  reviewed \cite{Podesta:2013}.} Although the nonlinear evolution of
kinetic \Alfven wave collisions remains to be solved rigorously, I
expect that, although the  resulting energy transfer
in the kinetic regime may differ quantitatively from that of MHD
\Alfven wave collisions, the qualitative picture of nonlinear energy
transfer by counterpropagating wave collisions persists.

Although the collisionless damping of the \Alfven waves is weak in the
inertial range at $k_\perp \rho_i \ll 1$, ion Landau damping peaks at
$k_\perp \rho_i \sim 1$ and is expected to transfer energy from the
turbulent electromagnetic fluctuations into non-thermal free energy in
velocity space through collisionless wave-particle interactions. The
resulting small-scale structure in the ion velocity distribution,
through a nonlinear phase mixing process, may undergo a dual-cascade
to smaller scales in both configuration space and velocity space,
termed the ion entropy cascade
\cite{Schekochihin:2009,Tatsuno:2009,Plunk:2010,Plunk:2011,Kawamori:2013}.
The entropy cascade enables efficient collisional thermalization of
the non-thermal free energy associated with fluctuations in the ion
distribution function even under conditions of arbitrarily weak
collisionality. In addition to ion Landau damping, electron Landau
damping becomes increasingly strong as the perpendicular wavenumber
increases within the dissipation range, $k_\perp \rho_i \gtrsim 1$.
Ultimately, the electron Landau damping becomes sufficiently strong to
dominate over the nonlinear energy transfer, terminating the cascade
with an exponential roll-off of the magnetic energy spectrum around
the scale of the electron Larmor radius, $k_\perp \rho_e \sim 1$
\cite{Alexandrova:2012,TenBarge:2013b}.

Based on such a dynamical understanding of \Alfvenic turbulence and
its collisionless damping, it is possible to construct predictive
models for the turbulent cascade of energy and its dissipation
\cite{Howes:2008b,Jiang:2009,Podesta:2010a,Howes:2011b,Schreiner:2014}.
Turbulent cascade models have been used to predict: the density
fluctuations in the solar corona and solar wind
\cite{Chandran:2009b,Chen:2013a}; the ratio of ion to electron heating
as a function of $\beta_i$ and $T_i/T_e$ \cite{Howes:2010d}; and the
proton-to-total turbulent heating ratio in the solar wind
\cite{Howes:2011c}. As well, such cascade models have proven
invaluable in the interpretation of kinetic numerical simulations of
weakly collisional plasma turbulence
\cite{Howes:2008a,Howes:2011b,TenBarge:2013b}.

%===================================================================
\subsection{Relation to Stochastic Heating}
Although it may not be immediately obvious, the model of stochastic
heating is entirely compatible with the dynamical model of anisotropic
\Alfvenic turbulence described above. Recent work has resulted in a
very well-developed model of stochastic ion heating in heliospheric
plasma turbulence
\cite{Chandran:2010a,Chandran:2010b,Chandran:2011,Bourouaine:2013}. A
key component of this model is a turbulent spectrum of \Alfven and
kinetic \Alfven waves, such as the \Alfvenic cascade described above.
If a particular ion has a Larmor radius\footnote{Note that this Larmor
  radius $\rho$ is \emph{not} the thermal ion Larmor radius $\rho_i =
  v_{ti}/\Omega_i$, but rather is the gyroradius for an ion with a
  particular perpendicular velocity $v_\perp$.} $\rho= v_\perp
/\Omega_i$, when the amplitude of the turbulent \Alfven and kinetic
\Alfven fluctuations at $k_\perp \rho \sim 1$ exceeds a certain
threshold, the ion's orbit becomes chaotic. The resulting stochastic
interaction of the ion with the time-varying electrostatic potential
associated with the turbulent fluctuations leads to a random walk of
the ion's energy. For a distribution of ions that monotonically
decreases with increasing energy, this leads to net damping of the
turbulent electromagnetic fluctuations and an increase in the
perpendicular temperature of the ion distribution.  This stochastic
ion heating mechanism is particularly effective under the low plasma
beta conditions relevant to the solar corona, with estimates that half
or more of the turbulent cascade power may be diverted into proton
heating at the scale of the proton thermal Larmor radius, $k_\perp
\rho_i \sim 1$ \cite{Chandran:2010a}. Thus, incoherent wave-particle
interactions may serve as an effective channel to dissipate plasma
turbulence.

In principle, stochastic heating could be incorporated into the
cascade models mentioned in \secref{sec:alfturb}, but to attempt such
an implementation raises several unanswered questions: (i) what
fraction of the large-scale turbulent cascade energy is stochastically
lost to the ions? (ii) how is this energy loss distributed over
different scales of the turbulent \Alfvenic fluctuations?, and (iii)
does this mechanism impact the nature of the fluctuations at smaller
scales that receive the remaining turbulent energy not lost
stochastically to the ions? Indeed, a recent observational study
finding an amplitude dependent drop in the magnetic energy spectrum at
the ion kinetic scale \cite{Bruno:2014b} may be evidence of turbulent
cascade energy lost stochastically to ions.

%===================================================================
\subsection{Dissipation Associated with Current Sheets}
Although the property that dissipation in plasma turbulence occurs
dominantly in small-scale current sheets is well established
\cite{Biskamp:1989,Spangler:1998,Spangler:1999,Biskamp:2000,Maron:2001,Merrifield:2005,Wan:2012,Karimabadi:2013,TenBarge:2013a,Wu:2013a,Zhdankin:2013},
the kinetic physical mechanism by which dissipation occurs in current
sheets has not been elucidated.  Indeed, stochastic ion heating
\cite{Parashar:2009,Markovskii:2011} and Landau damping
\cite{Loureiro:2013,TenBarge:2013a,Numata:2014} have both been
suggested as the physical process underlying the kinetic damping of
current sheets. In addition, parallel electric fields
\cite{Drake:2003,Pritchett:2004,Egedal:2008,Egedal:2009,Egedal:2010,Egedal:2012}
and Fermi acceleration \cite{Drake:2006}, both arising through the
process of magnetic reconnection, have also been proposed. It should
be emphasized, however, that even though magnetic reconnection often
springs to mind when current sheets are found, it has not yet been
firmly established that magnetic reconnection plays a significant role
in the dissipation of energy in three-dimensional plasma turbulence in
the solar wind. The topics of the connection between magnetic
reconnection and turbulence and the role played by current sheets in
the dissipation of turbulence are central to several reviews in this
issue \cite{Goldstein:2015,Lazarian:2015,Matthaeus:2015}.

One important question is whether the current sheets inferred from
solar wind observations are all dynamically generated by the
turbulence itself \cite{Boldyrev:2011,Zhdankin:2012}, or whether some
fraction of the current sheets merely represent advected flux tube
boundaries \cite{Borovsky:2008,Borovsky:2010,Miao:2011,Arnold:2013}.
I have highlighted in \secref{sec:curr} recent research that explains,
from first-principles, the development of current sheets in plasma
turbulence as a natural consequence of the nonlinear energy transfer
caused by strong \Alfven wave collisions \cite{Howes:2014d}. If this
mechanism operates in the solar wind, then at least some of the
observed current sheets consist of a constructively interfering sum of
counterpropagating \Alfven waves. This finding provides new insight
into the underlying nature of dynamically generated current sheets in
plasma turbulence, and the damping of the associated turbulent
electromagnetic fluctuations may indeed be dominated by collisionless
wave-particle interactions via the Landau resonance, as previously
suggested by a recent analysis of gyrokinetic numerical simulations
\cite{TenBarge:2013a}. If these conjectures are correct, then
\emph{the fact that the heating is concentrated in current sheets is
  merely a consequence of the turbulent nonlinear dynamics but does
  not determine the mechanism of dissipation}. This idea does not
conflict with results reported elsewhere in this issue suggesting that
intermittency develops identically in both ideal and resistive MHD
simulations \cite{Matthaeus:2015}.

A statement that is likely to prove very controversial is that
intermittency, although a well established characteristic of plasma
turbulence, may not be critically important for the prediction of the
plasma heating due to the dissipation of turbulent fluctuations,
particularly if coherent (or possibly even stochastic) wave-particle
interactions dominate the dissipation. This directly contradicts the
conclusion of Matthaeus \emph{et al.} \cite{Matthaeus:2015} that
``recent studies of kinetic dissipation of the turbulent cascade
suggest that coherent structures and associated nonuniform dissipation
play a very important and possibly dominant role in the termination of
the cascade and the effectively irreversible conversion of fluid
macroscopic energy into microscopic random motions.''  In short, the
viewpoint proposed here is that current sheets are not a cause of
turbulent dissipation, but are merely a consequence of the nonlinear
dynamics underlying the turbulent energy cascade.  If this contentious
statement is correct, it represents good news for the endeavor to
develop predictive models of plasma heating arising from the
dissipation of weakly collisional plasma turbulence.  Simple cascade
models would remain a valid means of predicting the differential
heating of the plasma species since the collisionless damping would
depend only on the energy content in different wavevectors, while the
phases that lead to intermittency would not come into play.

%===================================================================
\section{Dynamical Model of Plasma Turbulence}
\begin{figure}[p]
\centering\hbox{\resizebox{5.25in}{!}{\includegraphics{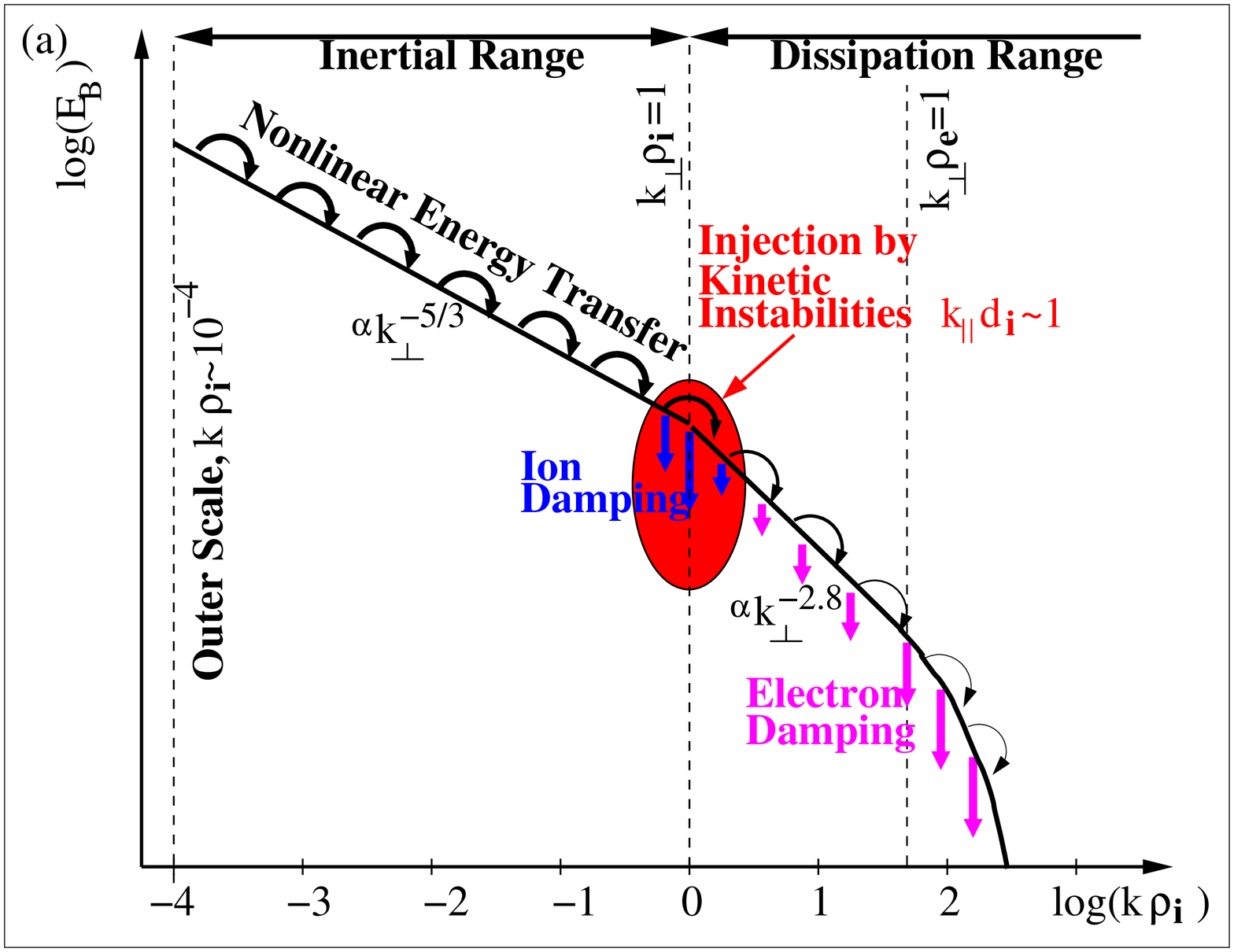}}}
\centering\hbox{\resizebox{5.25in}{!}{\includegraphics{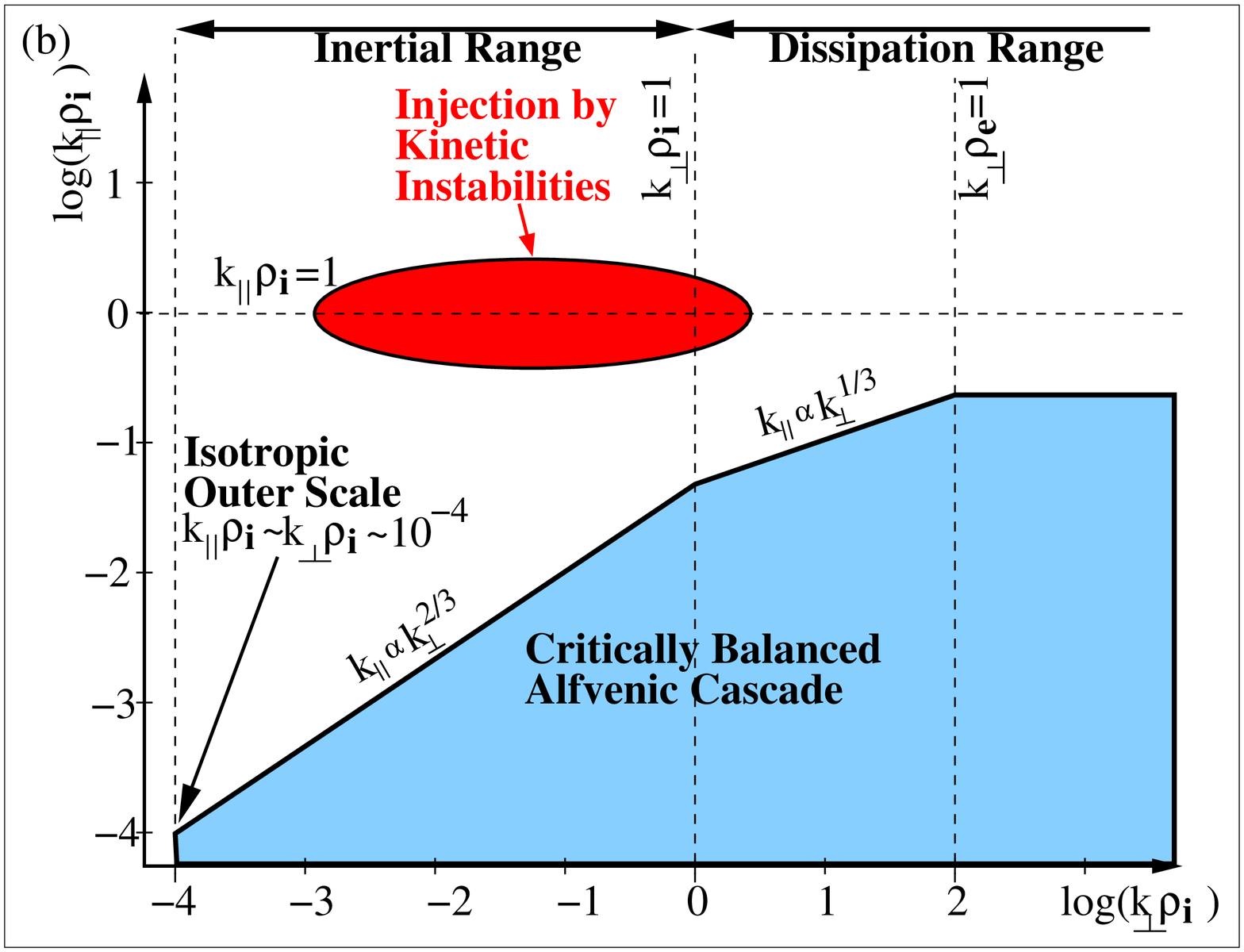}}}
\caption{(a) Diagram of the magnetic energy spectrum in plasma
  turbulence, including the injection of energy by kinetic
  instabilities. (b) Anisotropic distribution of power in
  $(k_\perp,k_\parallel)$ wavevector space due to both the cascade of
  energy from large scales (shaded) and the injection of energy by kinetic
  instabilities (ellipse).
\vspace*{-10pt} }
\label{fig:model}
\end{figure}

Here I present a brief summary of the first dynamical model of plasma
turbulence that combines the physics of \Alfven waves with the
self-consistent development of current sheets, illustrated by the
diagrams of the magnetic energy spectrum and wavevector anisotropy in
\figref{fig:model}.

Isotropic fluctuations at the outer scale of the turbulent inertial
range in the near-Earth solar wind, typically at $k \rho_i \sim
10^{-4}$, consist primarily of incompressible \Alfven waves with a
small admixture of kinetic slow wave fluctuations (not depicted).
Nonlinear energy transfer from large to small scales through the
inertial range is governed by the physics of \Alfven wave collisions.
The theory of the critically balanced \Alfvenic turbulence
\cite{Goldreich:1995} predicts a magnetic energy spectrum $E_B \propto
k_\perp^{-5/3}$, where the turbulent fluctuations become increasingly more
anisotropic as the cascade progresses to smaller scales, following a
scale-dependent anisotropy $k_\parallel \propto k_\perp^{2/3}$ in the
inertial range. Note that a refined version of critically balanced
turbulence that accounts for the dynamic alignment of the turbulent
fluctuations \cite{Boldyrev:2006} predicts slightly different
scalings, $E_B \propto k_\perp^{-3/2}$ and $k_\parallel \propto
k_\perp^{1/2}$, and appears to be more consistent with the high
resolution numerical simulations of incompressible MHD turbulence
\cite{Perez:2012}.

At the perpendicular scale of the ion Larmor radius, $k_\perp \rho_i
\sim 1$, the turbulent cascade transitions from the inertial range
($k_\perp \rho_i \ll 1$), in which the nonlinear energy transfer is
mediated by \Alfven wave collisions, to the dissipation range ($k_\perp
\rho_i \gtrsim 1$), in which the nonlinear energy transfer is mediated
by kinetic \Alfven wave collisions. The dispersive nature of kinetic
\Alfven waves leads to a steepening of the magnetic energy spectrum
$E_B \propto k_\perp ^{-2.8}$, and an extension of the concept of
critical balance to this regime \cite{Howes:2008b} predicts an 
anisotropy  scaling as $k_\parallel \propto k_\perp ^{1/3}$.

The nonlinear energy transfer is governed by \Alfven and kinetic
\Alfven wave collisions, which transfer energy to \Alfvenic waves at
smaller scales. The particular phases and amplitudes of the
nonlinearly generated waves, determined by the mathematical form of
the nonlinear terms in the dynamical equations, leads to the
development of coherent structures through constructive interference
\cite{Howes:2014d}.  In plasma turbulence, these coherent structures
take the form of current sheets, and dissipation is found to be
concentrated within these current sheets.  But, since the current
sheets themselves are primarily made up of counterpropagating \Alfven
and kinetic \Alfven waves, these turbulent electromagnetic
fluctuations will damp by collisionless wave-particle interactions via
the Landau resonance. Ion Landau damping peaks at scales $k_\perp
\rho_i \sim 1$, and electron Landau damping begins to be significant
at $k_\perp \rho_i \gtrsim 1$ and becomes increasingly strong as the
wavenumber increases.  At the electron scales $k_\perp \rho_e \sim 1$,
electron Landau damping overwhelms the nonlinear energy
transfer and the cascade is terminated, producing an exponential
roll-off in the magnetic energy spectrum. Ultimately, the energy
removed from the turbulent fluctuations by collisionless wave-particle
interactions will ultimately be converted to plasma heat by
collisions.  This process is facilitated by an entropy cascade (not
shown) that mediates the transfer of nonthermal structure to
sufficiently small scales in velocity space that weak collisions can
thermalize the energy \cite{Schekochihin:2009}.

In addition to the physics of the \Alfvenic turbulent cascade of
energy from large scales that is ultimately terminated at electron
scales, kinetic instabilities can inject energy into fluctuations at
scales\footnote{Note that $d_i = \rho_i/\sqrt{\beta_i}$, so for the
  typical value $\beta_i \sim 1$ in the near-Earth solar wind, $d_i
  \sim \rho_i$.} $k_\parallel d_i \sim 1$. An important, but generally
under-appreciated, property of these instability-driven fluctuations
is that they occupy a distinct region of wavevector space from the
anisotropic cascade of energy from large scales. However, since
single-point spacecraft measurements cannot uniquely separate
$k_\parallel$ from $k_\perp$, these instability-driven fluctuations
will appear at $k d_i \sim k \rho_i \sim 1$ in the magnetic energy
spectrum, making them quite difficult to distinguish.

%===================================================================
\section{Conclusion}
An improved understanding of turbulence in the weakly collisional
solar wind will impact our ability not only to predict physical
processes within the heliosphere but also to illuminate complex
astrophysical phenomena in remote regions of the universe. The linear
response due to magnetic tension in a magnetized plasma, a physical
property absent in turbulent hydrodynamical flows, provides an
important foothold to understand the dynamics of plasma turbulence. To
unravel the physics of the nonlinear energy transfer to small scales,
the kinetic mechanisms of dissipation of the turbulent fluctuations,
and the resulting plasma heating in kinetic plasma turbulence, I
contend that we must step beyond the usual statistical treatments of
turbulence, adopting instead a dynamical approach.  It is the linear
and nonlinear dynamics of \Alfven waves that are responsible, at a
very fundamental level, for some of the key qualitative features of
plasma turbulence that distinguish it from hydrodynamic turbulence,
including the anisotropic cascade of energy and the development of
current sheets at small scales.  The ultimate goal is to create a
predictive theory of plasma turbulence.  Only a thorough understanding
of the turbulent plasma dynamics will enable the development of a
simplified theoretical framework upon which predictive models of
plasma turbulence and its effect on energy transport and plasma
heating can be constructed.

The arguments presented here support a simplified perspective on the
nature of turbulence in a weakly collisional plasma.  The nonlinear
interactions responsible for the turbulent cascade of energy and the
formation of current sheets are essentially fluid in nature, and can
be understood in terms of nonlinear wave-wave interactions.  On the
other hand, the collisionless damping of the turbulent fluctuations
and the energy injection by kinetic instabilities are essentially
kinetic in nature, and can be understood in terms of linear
collisionless wave-particle interactions.

From this perspective, it is easy to understand why reduced models,
such as incompressible MHD and reduced MHD, are valuable tools in the
study of plasma turbulence. The simplicity of such reduced models is
that they satisfy a number of exact constraints that facilitate the
development of an intuitive understanding of the turbulent
dynamics. Although these constraints do not strictly hold under more
general plasma conditions, often the behavior observed in the
simplified system persists, at least to lowest order, in the more
general system.  For example, we employ here our understanding of the
nonlinear energy transfer between counterpropagating \Alfven waves in
incompressible MHD to argue that the anisotropic cascade of energy
ubiquitously observed in magnetized plasma turbulence is due to the
facts that only counterpropagating \Alfven waves interact nonlinearly
and that the nonlinear term is greatest for interactions between
perpendicularly polarized \Alfven waves.  Although the constraint
that only counterpropagating waves interact ceases to hold strictly
when compressibility or kinetic effects (dispersion due to finite
Larmor radius effects) are introduced, the anisotropy of the turbulent
cascade persists.

The separation of essentially fluid versus essentially kinetic
properties of kinetic plasma turbulence enables further
simplifications.  Although non-Maxwellian particle velocity
distributions are widely found in the solar wind and in kinetic
numerical simulations, the departures from a Maxwellian equilibrium
(bi-Maxwellian, kappa, or multi-component with core/halo structure) do
not affect the electromagnetic dynamics of the turbulence very much
\cite{Klein:2015a}.  Since the nonlinear energy transfer and
development of current sheets in plasma turbulence are due to
wave-wave interactions, they depend only on the first moments of the
distribution functions, and are therefore not very sensitive to the
particular form of the distribution functions.  The physical
mechanisms of dissipation and the injection of energy by kinetic
instabilities, on the other hand, which depend on growth or damping
rates due to collisionless wave-particle interactions, may depend
sensitively on the form of the distribution functions.  For example, a
kappa distribution, having a much larger population of particles at
high energy with respect to a Maxwellian, may experience significantly
enhanced rates of collisionless damping of waves that are resonant
with these high energy particles.  This viewpoint proposed here
strongly contradicts the recent suggestion from a kinetic study of
plasma turbulence \cite{Servidio:2014c} that ``it seems increasingly
clear that significant kinetic effects including heating have strong
association with coherent structures and with the turbulent cascade
that produces intermittency.''

Looking toward the future, to identify definitively the kinetic
mechanism responsible for the dissipation of solar wind turbulence, it
is likely that it will be necessary to step beyond the analysis of the
turbulent electromagnetic fields alone, delving much more deeply into
the dynamic effect of turbulent fluctuations and their dissipation on
the particle velocity distributions.  These velocity distributions are
already measured directly by instruments on existing spacecraft
missions. And, on upcoming missions, such as \emph{MMS} and
\emph{Solar Probe Plus}, with a combination of higher cadence and
better resolution than previous missions, the distribution function
data represent an important untapped potential source for discovery
science. A new frontier will be the exploration of the dynamics of the
perturbations to the distribution functions associated with the
turbulent fluctuations and their kinetic dissipation, where kinetic
plasma turbulence simulations will be essential to exploit fully the
potential of \emph{MMS} and \emph{Solar Probe Plus}.

\section*{Acknowledgment}
I would like to thank Jason TenBarge and Kris Klein for their
invaluable feedback and contributions. Financial support has been
provided by NSF grant PHY-10033446, NSF CAREER Award AGS-1054061, and
NASA grant NNX10AC91G.

%%%%%%%%%% Insert bibliography here %%%%%%%%%%%%%%
%\bibliographystyle{is-unsrt}
%\bibliography{abbrev2,solwind,gyro,turbulence,gen_plasma,parametric,instability,reconnection,space,corona,lab,ptrsa}

\end{document}